\documentclass{article}[12pt] 
\usepackage{ amsmath, amsthm, amssymb } 
\usepackage{color} 
\usepackage{cite}

\newcommand{\be} { \begin{equation} } 
\newcommand{\ee} { \end{equation} } 

\newcommand{\labbel}[1] { \label{#1} } 
\newcommand{\Du}{ D\left(u,\frac{d}{du}\right) } 
\newcommand{\Duu}{ D_1\left(u,\frac{d}{du}\right) } 
\newcommand{\rb}{ \sqrt{b} }
\newcommand{\ru}{ \sqrt{u} }
\newcommand{\ruz}{ \sqrt{u_0} }

\newcommand{\rR} {\sqrt{R_4(u,b)}}

\newcommand{\EG}{{\rm EG}}
\newcommand{\EI}{{\rm EI}}
\newcommand{\EJ}{{\rm EJ}}

\newcommand{\ie}{{\it i.e. }}

\newcommand{\Ima}{{\rm Im}}

\newcommand{\Li}{{\rm Li}}

\newcommand{\sunrise}[1]{ 
\mbox{\parbox{2.5cm}{\hspace{0.25cm} 
\begin{picture}(2,1) 
\thicklines 
\put(0.3,0.5){\vector(1,0){0.1}} 
\put(0.5,0.5){\line(1,0){1}} 
\put(0,0.5){\line(1,0){0.5}} 
\put(1.5,0.5){\line(1,0){0.5}} 
\put(1,0.5){\circle{2}} 
\put(0.85,1.12){$m$} 
\put(0.85,0.60){$m$} 
\put(0.85,0.08){$m$} 
\put(0.25,0.7){\makebox(0,0)[b]{$#1$}} 
\end{picture} 
}}}

\begin{document} 
\setlength{\unitlength}{1.3cm} 
\begin{titlepage}
\vspace*{-1cm}
\begin{flushright} TTP17-038 \end{flushright}                                
\vskip 3.5cm
\begin{center}
\boldmath
 
{\Large\bf An Elliptic Generalization of Multiple Polylogarithms} 
\unboldmath
\vskip 1.cm
{\large Ettore Remiddi}$^{a,}$
\footnote{{\tt e-mail: ettore.remiddi@bo.infn.it}}
{\large and Lorenzo Tancredi}$^{b,}$
\footnote{{\tt e-mail: lorenzo.tancredi@kit.edu}} 
\vskip .7cm
{\it $^a$ DIFA, Universit\`a di Bologna and INFN, Sezione di Bologna, 
I-40126 Bologna, Italy } \\
{\it $^b$ Institute for Theoretical Particle Physics, KIT, 76128 Karlsruhe, 
Germany } 
\end{center}
\vskip 2.6cm

\begin{abstract}
We introduce a class of functions which constitutes an obvious  
elliptic generalization of multiple polylogarithms. A subset of these functions
appears naturally in the $\epsilon$-expansion of the imaginary part
of the two-loop massive sunrise graph.
Building upon the well known properties of multiple polylogarithms, we
associate a concept of weight to these functions and show that
this weight can be
lowered by the action of a suitable differential operator. 
We then
show how properties and relations among these functions 
can be studied  bottom-up starting from lower weights.

\end{abstract} 
\vfill 
\end{titlepage} 
\newpage 
\section{Introduction} \labbel{sec:intro} \setcounter{equation}{0} 
\numberwithin{equation}{section} 
The generalized 
polylogarithms~\cite{Goncharov,Remiddi:1999ew,Gehrmann:2000zt,Vollinga:2004sn} 
(also called Goncharov functions) are 
of common use in the evaluation of Feynman graph amplitudes, especially 
in the differential equation approach. As it is well known, however, 
they are not enough to span the 
full set of functions required to evaluate two-loop Feynman integrals. 
The obvious obstruction comes from Feynman graphs that fulfil irreducible 
second- (or higher-) order differential equations, of which the most notable 
example is indeed the massive two-loop sunrise graph. In spite of the long 
efforts and the vast literature produced on the 
subject~\cite{Broadhurst:1987ei,Bauberger:1994by,Bauberger:1994hx,
Caffo:1998du,Laporta:2004rb,Bloch:2013tra,Remiddi:2013joa,
Adams:2013nia,Adams:2014vja,Adams:2015gva,Adams:2015ydq,Remiddi:2016gno,
Adams:2016xah,Bonciani:2016qxi,Bloch:2016izu,Passarino:2016zcd,vonManteuffel:2017hms,Adams:2017ejb,Ablinger:2017bjx}, the way to further generalize the polylogarithms 
to accomodate this case remains topic of discussion, with a fascinating
crosstalk between particle physics, mathematics and string theory, see for 
example~\cite{Brown:MEPLs,Broedel:2015hia}.
\par 
In this paper we introduce and discuss an elliptic generalization
of multiple polylogarithms, 
referred to in this Introduction with the name $ \EG^{[n]} $ for short, 
(more refined notations will be used in the next Sections), 
defined starting from an integral representation of the form 
\be \EG^{[n]}(k,u) = \int_{b_i}^{b_j} \frac{db\, b^k}{\rR}\ g^{[n]}(u,b) \ , 
                                                          \nonumber \ee 
where 
$ R_4(u,b) $ is the fourth order polynomial in $b$ 
\be R_4(u,b) = b(b-4m^2)[(\ru-m)^2-b][(\ru+m)^2-b] \ ,   \nonumber \ee 
$ (b_i, b_j) $ is any pair of the 4 roots of $ R_4(u,b) $, namely 
$ b_1=0, b_2=4m^2, b_3=(W-m)^2 $ and $ b_4=(W+m)^2 $ 
and $ g^{[n]}(u,b) $ is a generalized polylogarithm in $ b $ of degree 
$ n $ with ``alphabet" corresponding to the above 4 roots $ b_i $ 
(for simplicity we consider mainly $ u $ real and in the range 
$ 9m^2 < u < \infty $, but the continuation to other values of $ u $ 
is almost obvious). \par 
The $ \EG^{[n]}(k,u) $ are generalizations of the integrals
\begin{align} I(k,u) &= \int_{4m^2}^{(\ru-m)^2} 
                    \frac{db\, b^k}{\sqrt{R_4(u,b)}} \,.
\nonumber 
\end{align} 
The latter can be expressed in terms of two
independent master integrals, say $I_0(u)$ and $I_2(u)$, 
which are simple suitable linear combinations of the $I(k,u)$ and
correspond to $K(x)$ and $E(x)$, the complete elliptic integrals
of first and second kind respectively 
(hence the {\it elliptic} terminology for the new functions); one has for example
\begin{align}
     I_0(u) & =\frac{2}{\sqrt{(\ru+3m)(\ru-m)^3}} 
        \ K\left(\frac{(\ru-3m)(\ru+m)^3}{(\ru+3m)(\ru-m)^3}\right) \ .
\nonumber 
\end{align} 
Moreover, up to an 
inessential numerical factor, $ I_0(u) $ is the phase space of three 
particles of equal mass $ m $ at energy $ \ru $ in $ d=2 $ dimensions, 
(see Section~\ref{sec:sun}) while the newly introduced $ \EG^{[n]}(k,u) $ 
are obvious generalizations of the terms which arise when expanding the 
$d$-dimensional 3-body phase space in powers of $(d-2)$. 
A subset of the functions $\EG^{[n]}(k,u) $ is therefore naturally appearing in the
$(d-2)$-expansion of the imaginary part of the two-loop massive
sunrise graph.
\par 

$ I_0(u) $ and $I_2(u)$ satisfy a {\it homogeneous} two by two system of 
linear differential equations in $u$  (see Eq.~\eqref{eq:IJ2b}), which can be written as
\be \frac{d}{du} \left( \begin{array}{c} I_0(u) \\ I_2(u) \end{array} \right) = 
\left( \begin{array}{cc} B_{0,0}(u) & B_{0,2}(u) \\  B_{2,0}(u) & B_{2,2}(u) \end{array} \right) 
\left( \begin{array}{c} I_0(u) \\ I_2(u) \end{array} \right) 
= B(u) \left( \begin{array}{c} I_0(u) \\ I_2(u) \end{array} \right)  \ , 
 \nonumber \ee 
where the matrix elements of the $ 2 \times 2 $ matrix $B(u)$, given in 
Eq.s(\ref{eq:udI0u},\ref{eq:udI2u}), contain rational coefficients and 
poles at $ u $ equal to $ 0, m^2, 9m^2$. 
\par 
At variance with the $I_k(u)$, the study of the functions $ \EG^{[n]}(k,u) $,
for every value of $n$, requires the introduction of three master integrals
instead of two. We consider then
three new functions $ \EG_k^{[n]}(u)$, $k=0,1,2 $, which are again
simple linear combinations of the above $ \EG^{[n]}(k,u) $, 
(see for instance Eq.~\eqref{eq:Inu}), and we find that they satisfy an {\it inhomogeneous} 
system of differential equations of the form 
\begin{align} 
&\frac{d}{du} \left( \begin{array}{c} \EG_{0}^{[n]}(u) \\  \EG_{2}^{[n]}(u) \end{array} \right) =
B(u) \left( \begin{array}{c}  \EG_{0}^{[n]}(u)  \\   \EG_{2}^{[n]}(u)  \end{array} \right) 
+  \sum_{k=0,1,2}   \left( \begin{array}{c} 
                R_{0,k}^{[n-1]}(u) \EG_{k}^{[n-1]}(u)   \\   
                R_{2,k}^{[n-1]}(u) \EG_{k}^{[n-1]}(u) \end{array} \right) \nonumber  \\ 
&\frac{d}{du} \EG_1^{[n]}(u) = 
  \sum_{k=0,1,2} 
                R_{1,k}^{[n-1]}(u) \EG_{k}^{[n-1]}(u)\,, \labbel{eq:EpolMatrixdef} 
 \end{align}

where the matrix $B(u)$ is the same as in the previous 
homogeneous equation for the $ I_k(u) $, while the coefficients 
$ R_{k,k'}^{[j]}(u) $ of the inhomogeneous terms consist again in general of 
rational expressions in $ u $ with poles at $ u $ equal to $ 0, m^2, 9m^2 $.
As we will see, $\EG_k^{[0]}(u) = I_k(u)$, and one finds in particular that
$\EG_1^{[0]}(u)$ is a constant, which explains why for $n=0$ only
two independent functions are needed instead of three.

\par
Note the presence, in the {\it r.h.s.} of the above equations, of functions 
of the same family $ \EG_k^{[n]}(u) $, but 
with {\it lower} values of the index $ n $.
In particular, for any given $n$, we find a two by two system of coupled 
differential equations, plus a third, simpler, decoupled linear differential equation.
This suggests indeed that we can tentatively
associate a weight $ n $ to the functions $ \EG_k^{[n]}(u) $ with respect to
the action of a {\it three by three matricial operator}. The latter, though,
clearly factorises into
the two by two operator 
$ ( -B(u) + d/du ) $, which directly lowers the weight of $ \EG_0^{[n]}(u) $ 
and $ \EG_2^{[n]}(u) $,
and the simple first order differential operator, $d/du$, 
which lowers the weight of $\EG_1^{[n]}(u)$
(similarly to what happens with the Goncharov polylogarithms).
Such a generalized weight will be called 
E-weight and the functions $ \EG_{k=0,1,2}^{[n]}(u) $ also referred to as 
E-polylogarithms. 
\par 

As the pair of functions $ I_k(u) $, in particular, is annihilated by 
the operator $ ( -B(u) + d/du ) $, the two functions $ I_k(u) $ can 
be considered E-polylogarithms of E-weight equal zero. 
Similarly, at E-weight zero the third function, say $I_1(u)$,
is a constant and is therefore 
annihilated by $d/du$.
\par 

As we will see, it can be useful to rewrite the homogeneous first order system 
for $ I_0(u), I_2(u) $ as a homogeneous second order differential equation 
for $I_0(u)$ only; when that is done, one obtains 
\be D\left( u,\frac{d}{du} \right) I_0(u) = 0\ , \nonumber \ee 
where $ D(u,d/du) $, Eq.\eqref{eq:Du}, is a suitable second order 
differential operator. Acting similarly on the functions 
$ \EG_k^{[n]}(u) $ we find
\be D\left( u,\frac{d}{du} \right)\ \EG_0^{[n]}(u) = 
          \sum_{k=0,1,2}\, r_k^{[n-1]}(u) \EG_k^{[n-1]}(u)  
                            +   \sum_{k=0,1,2} r_k^{[n-2]}(u) \EG_k^{[n-2]}(u) 
                     \ , \labbel{eq:EpolDudef} \ee 
where $ D(u,d/du) $ is the same differential operator appearing in 
the second order equation satisfied by $ I_0(u) $, while 
the coefficients $ r_k^{[j]}(u) $ are also rational expressions in $ u $, 
with poles at $ u $ equal to $ 0, m^2, 9m^2 $. 
As we can see, in the {\it r.h.s} of the above equation we find 
E-polylogarithms of weight $n-1$ and $n-2$.
We can therefore also say that 
a function $ \EG_0^{[n]}(u) $ satisfying the above equations is an 
E-polylogarithm of E-weight $ n $ under the 
action of the scalar second order differential operator $ D(u,d/du) $. 
That confirms, of course, that $ I_0(u) $, being annihilated by $ D(u,d/du) $ 
has E-weight equal $ 0 $. 
Alternatively, one could also derive a different second order differential
operator, say $D_2(u,d/du)$, such that
$$D_2\left( \frac{d}{du}, u\right) I_2(u) = 0\,,$$
and the discussion would apply in the very same way.
\par 
In the course of the paper we will also encounter repeated integrations 
of rational factors times for instance the function $ I_0(u) $; 
in this picture, they constitute a simple subset 
of E-polylogarithms, and we will refer to them for simplicity as 
$E_0$-polylogarithms, 
see Section~\ref{sec:E1pol}.
\par
The equations for the $ \EG_k^{[n]}(u) $ can further be solved 
by using the Euler method, which provides representations for the 
$ \EG_k^{[n]}(u) $ as suitable integrals 
involving the solutions of the homogeneous equation, \ie the function 
$ I_0(u) $ above with the accompanying function $ J_0(u) $, Eq.\eqref{eq:J0u}, 
and the inhomogeneous term, providing interesting relations between the 
$ \EG_k^{[n]}(u) $ and the repeated integrations of products of the 
$ I_0(u), J_0(u) $ and the usual (poly)logarithms of $ u $. 
An example of such relations is
\be \int_{4m^2}^{(W-m)^2} \kern-15pt \frac{db}{\sqrt{R_4(u,b)}} 
    \ \ln{b}  = \frac{2}{3}\ln(u-m^2)\ I_0(u)  \nonumber \ee 
(where we have written $ \ln(b/m^2) $, for simplicity, as $ \ln{b} $ ). 
\par 
Let us recall, indeed, that one of the musts of an analytic calculation is 
to discuss as deeply and explicitly as possible the identities which might 
hold between the various functions introduced in the calculation. This 
allows one to write the result in a compact form and to understand whether 
two apparently different formulas are indeed different or equal.
\par
The rest of the paper is organized as follows: in Section~\ref{sec:Thebeg}
we study the (well known) functions $I(k,u)$ and reduce them to three master 
integrals using integration-by-parts. We then show that one of the 
three masters is not linearly independent and show how to derive 
a two by two system of  differential equations for the two masters 
$I_0(u)$ and $I_2(u)$, and their accompanying functions 
$J_0(u)$ and $J_2(u)$. In Section~\ref{sec:E1pol} we study a first simple 
class
of functions obtained by repeated
integrations of products of one of the $I_k(u)$ or $J_k(u)$ 
times rational factors. As the complexity of these functions is decreased by
differentiation, these functions can be given a simple concept
of weight, similar to that of multiple polylogarithms. We call these functions
$E_0$-polylogarithms and their weight $E_0$-weight.
Then in Section~\ref{sec:FF} we study the first example of E-polylogarithm
at E-weight one and show how it can be rewritten as the product of
a logarithm and the E-weight zero function $I_0(u)$.
Similar relations are derived for all E-weight one functions in Section~\ref{sec:Eweight1all}.
We extend then our study to higher weights in Section~\ref{sec:Eweight2} and find
explicit relations to simplify E-polylogarithms at E-weight 2.
Finally we use our results to give a compact representation of the imaginary
part of the twol-loop massive sunrise up to order $\epsilon^2$ in Section~\ref{sec:sun}.
We then draw our conclusions and outlook in Section~\ref{sec:conc}.

\section{The beginning} 
\labbel{sec:Thebeg} \setcounter{equation}{0} 
\numberwithin{equation}{section} 
In this Section we will start by recalling some known 
results, which will be generalized in the rest of this paper. 
To begin with, for $ 9m^2 < u < \infty $ we consider the (real) 
function 
\be I_0(u) = \int_{4m^2}^{(W-m)^2} 
                    \frac{db}{\sqrt{R_4(u,b)}} \ , \labbel{eq:I0u} \ee 
where $ R_4(u,b) $ is the fourth order polynomial in $b$ 
\begin{align} R_4(u,b) &= b(b-4m^2)\left[(u-b-m^2)^2-4m^2b\right] 
                                                           \nonumber\\ 
             &= b(b-4m^2)[(W-m)^2-b][(W+m)^2-b]            \nonumber\\ 
             &= b(b-4m^2)[u-(\sqrt{b}+m)^2][u-(\sqrt{b}-m)^2]  \ , 
\labbel{eq:R4} \end{align} 
with 
\be W = \sqrt{u} \ .  \labbel{defW} \ee  

Eq.~\eqref{eq:I0u} corresponds, up to a multiplicative constant, to the 
imaginary part of the equal mass sunrise amplitude in $ d=2 $ dimensions. 
For convenience of later use, let us recall that its value in the 
$ u \to 9m^2 $ limit is 
\be \lim_{u\to 9m^2} I_0(u)= \frac{\sqrt{3}}{12m^2}\pi \ , 
\labbel{eq:I0at9} \ee 
an almost elementary result which can be easily obtained by performing the 
change of integration variable 
\be b = 4m^2 + (W+m)(W-3m)t \labbel{eq:btot} \ee 
and then taking the $ W \to 3m $ limit. 
The direct calculation of the $b$-integral of Eq.(\ref{eq:I0u}) further gives 
\be I_0(u) =\frac{2}{\sqrt{(W+3m)(W-m)^3}} 
           \ K\left(\frac{(W-3m)(W+m)^3}{(W+3m)(W-m)^3}\right) \ , 
\labbel{eq:IeqK} \ee 
where $ K(x) $ is the complete elliptic integral of the first kind. \par 
As a first step, following previous works~\cite{Laporta:2004rb}, let us 
derive a (second order, homogeneous) differential equation for $ I_0(u) $. 
To that aim, we define the (related) functions 
\be I(b_i,b_j,n,u) = \int_{b_i}^{b_j}  \frac{db}{\sqrt{R_4(u,b)}} \ b^n \ , 
\labbel{eq:I(n,u)} \ee 
where $(b_i,b_j)$, as above, are any two (different) roots of the polynomial 
$ R_4(u,b) $ and $ n $ is an integer, so that 
Eq.(\ref{eq:I0u}) is recovered for $ b_i=4m^2$, $ b_j=(W-m)^2$ and 
$ n = 0 $. (A trivial remark: it is sufficient to consider for 
$b_i, b_j$ only the pairs of adjacent roots, as any other choice is a 
linear combination of them). 
\par 
One has the (obvious) identity 
\be \int_{b_i}^{b_j} db \frac{d}{db}\left( \sqrt{R_4(u,b})\ b^n 
                                    \right) = 0 \ ; \labbel{eq:dbbd} \ee 
by explicitly carrying out the $b$-derivative and by using the 
replacement $ \sqrt{R_4(u,b)} = R_4(u,b)/\sqrt{R_4(u,b)} $, one finds 
\begin{align} 
  (n+2)I(b_i,b_j,n+3,u)&-(2n+3)(u+3m^2)I(b_i,b_j,n+2,u) 
              +(n+1)(u+3m^2)^2I(b_i,b_j,n+1,u) \nonumber\\ 
   &-2(2n+1)m^2(u-m^2)^2 I(b_i,b_j,n,u) = 0 \ . \labbel{eq:idbn} 
\end{align} 
The above identity holds for any $n$ (except $ n \le -1 $ if $ b_i = 0 $ 
or $ b_j = 0 $); for integer positive $ n $, by using it (recursively, 
when needed) one can express any $ I(b_i,b_j,n,u) $, for $ n\ge 3 $, in terms 
of the 3 {\it master integrals} 
$ I(b_i,b_j,0,u), I(b_i,b_j,1,u), I(b_i,b_j,2,u). $ \par 
Consider now the (auxiliary) quantities 
\be Q(k,u) = \int_{b_i}^{b_j} db\ \sqrt{R_4(u,b})\ b^k \ , 
\labbel{eq:Qku} \ee 
with $ k = 0,1,2 $. By writing (again) 
$ \sqrt{R_4(u,b)} = R_4(u,b)/\sqrt{R_4(u,b)} $, 
and using Eq.s(\ref{eq:idbn}) one finds 
\be Q(k,u) = \sum_{n=0,1,2} c(k,n,u) I(b_i,b_j,n,u) \ , \labbel{eq:QkucI} \ee 
where the coefficients $ c(k,n,u) $ are (simple) polynomials in $ u $. 
From it one gets at once 
\be \frac{d}{du} Q(k,u) = \sum_{n=0,1,2} \left[ 
        \left( \frac{d}{du}c(k,n,u) \right)\ I(b_i,b_j,n,u) 
      + c(k,n,u) \left( \frac{d}{du} I(b_i,b_j,n,u) \right) \right] \ . 
\labbel{eq:ud1Qku} \ee 
But we can obtain the $u$-derivative of the $ Q(k,u) $ by differentiating 
directly the definition Eq.(\ref{eq:Qku}), obtaining 
\begin{align} \frac{d}{du} Q(k,u) &= \int_{b_i}^{b_j} 
          db\ \left( \frac{d}{du} \sqrt{R_4(u,b})\right) \ b^k \nonumber\\ 
  &= \int_{b_i}^{b_j}db\ 
     \frac{b(b-4m^2)(u-b-m^2)}{\sqrt{R_4(u,b)}} b^k \nonumber\\ 
  &= \sum_{n=0,1,2} d(k,n,u) I(b_i,b_j,n,u) \ , \labbel{eq:ud2Qku} 
\end{align} 
where Eq.s(\ref{eq:idbn}) were again used, and the $ d(k,n,u) $ are also 
(simple) polynomials in $ u $. \par 
By writing, for a given value of $ k $, that the {\it r.h.s.} of 
Eq.(\ref{eq:ud1Qku}) is equal to the {\it r.h.s.} of Eq.(\ref{eq:ud2Qku}) 
one obtains a linear (homogeneous) equation 
expressing the $u$-derivatives of the three {\it master integrals} 
$ I(b_i,b_j,n,u) $, with $ n=0,1,2 $, in terms of the same three 
{\it master integrals}. One can then take three such equations, 
corresponding to three different values of $k$, say $k=0,1,2$ for 
definiteness, and solve them for the three derivatives. The result 
can be written as 
\be \frac{d}{du} \left(\begin{array}{c} 
               I(b_i,b_j,0,u) \\ 
               I(b_i,b_j,1,u) \\ 
               I(b_i,b_j,2,u) \end{array}\right) = 
    \left(\begin{array}{ccc} 
          C_{00}(u) & C_{01}(u) & C_{02}(u) \\ 
          C_{10}(u) & C_{11}(u) & C_{12}(u) \\ 
          C_{20}(u) & C_{21}(u) & C_{22}(u) \end{array}\right) 
\left(\begin{array}{c} 
               I(b_i,b_j,0,u) \\ 
               I(b_i,b_j,1,u) \\ 
               I(b_i,b_j,2,u) \end{array}\right) \ , \labbel{eq:udI(n)} 
\ee 
where 
\begin{align} 
    C_{00}(u) &= - \frac{1}{3u} - \frac{2}{3(u-9m^2)} \ , \nonumber\\ 
    C_{01}(u) &= \frac{1}{m^2}\left( \frac{1}{2u} - \frac{3}{4(u-m^2)} 
                      + \frac{1}{4(u-9m^2)} \right)   \ , \nonumber\\ 
    C_{02}(u) &= \frac{1}{m^4}\left( - \frac{1}{6u} + \frac{3}{16(u-m^2)} 
                - \frac{1}{48(u-9m^2)} \right) \ ,    \labbel{eq:udI(0)} 
\end{align} 

\begin{align} 
    C_{10}(u) &= m^2\left(- \frac{1}{3u} - \frac{8}{3(u-9m^2)} \right) 
                                                           \ , \nonumber\\ 
    C_{11}(u) &=  \frac{1}{2u} - \frac{1}{u-m^2} + \frac{1}{u-9m^2} 
                                                           \ , \nonumber\\ 
    C_{12}(u) &= \frac{1}{m^2}\left( - \frac{1}{6u} + \frac{1}{4(u-m^2)} 
               -  \frac{1}{12(u-9m^2)} \right)  \ , \labbel{eq:udI(1)} 
\end{align} 

\begin{align} 
    C_{20}(u) &= m^2\left(- 1 - \frac{m^2}{3u} -\frac{32m^2}{3(u-9m^2)} 
                                        \right) \ , \nonumber\\ 
    C_{21}(u) &= \frac{1}{2} + \frac{m^2}{2u} - \frac{4m^2}{u-m^2} 
                      + \frac{4m^2}{u-9m^2}     \ , \nonumber\\ 
    C_{22}(u) &= - \frac{1}{6u} + \frac{1}{u-m^2} 
                       -  \frac{1}{3(u-9m^2)} \ . \labbel{eq:udI(2)} 
\end{align} 
Eq.(\ref{eq:udI(n)}) is a linear homogeneous system of three first order 
differential equations 
for the three {\it master integrals} $ I(b_i,b_j,k,u) $, $ k=0,1,2 $. 
\par 
Quite in general, a three by three first order system is equivalent to a 
{\it third order} differential equation for one of the three functions, 
say for instance $ I(b_i,b_j,0,u) $; but we are looking for a 
{\it second order} equation. It is indeed known 
(see Appendix \ref{sec:AppX}) 
that 
one of the equations can be decoupled from the other two. To that aim, we 
introduce a new basis of {\it master integrals} according to the definitions 
\begin{align} 
I_0(b_i,b_j,u) &= I(b_i,b_j,0,u) \nonumber\\ 
     &= \int_{b_i}^{b_j}db\ \frac{db}{\sqrt{R_4(u,b)}} \ , \nonumber\\ 
I_1(b_i,b_j,u) &= I(b_i,b_j,1,u) - \frac{u+3m^2}{3} I(b_i,b_j,0,u) 
                                                              \nonumber\\ 
     &= \int_{b_i}^{b_j}db\ \frac{db}{\sqrt{R_4(u,b)}} 
        \left(b - \frac{u+3m^2}{3} \right) \ , \nonumber\\ 
I_2(b_i,b_j,u) &= I(b_i,b_j,2,u) - (u+3m^2) I(b_i,b_j,1,u) 
                         + \frac{(u+3m^2)^2}{3} I(b_i,b_j,0,u) \nonumber\\ 
     &= \int_{b_i}^{b_j}db\ \frac{db}{\sqrt{R_4(u,b)}} 
        \left( b^2 - (u+3m^2)b + \frac{(u+3m^2)^2}{3} \right) \ . 
                                                           \labbel{eq:Inu} 
\end{align} 
In terms of the functions of the new basis the system splits into a 
very simple equation involving only $ I_1(b_i,b_j,u) $, 
\begin{align} \frac{d}{du}I_1(b_i,b_j,u) &= 
         \frac{d}{du}\int_{b_i}^{b_j}db\ \frac{db}{\sqrt{R_4(u,b)}}
         \left(b - \frac{u+3m^2}{3} \right) \nonumber\\
        &= 0 \ ,        \labbel{eq:udI1u} \end{align} 
a result already noted by A.Sabry in his (1962) paper~\cite{Sabry} 
(see also Appendix~\ref{sec:AppX}), and in a two by two first order system 
for the other two functions $ I_0(b_i,b_j,u), I_2(b_i,b_j,u) $ 
\be \frac{d}{du} \left(\begin{array}{c} 
               I_0(b_i,b_j,u) \\ 
               I_2(b_i,b_j,u) \end{array}\right) = 
    \left(\begin{array}{cc} 
          B_{00}(u) & B_{02}(u) \\ 
          B_{20}(u) & B_{22}(u) \end{array}\right) 
\left(\begin{array}{c} 
               I_0(b_i,b_j,u) \\ 
               I_2(b_i,b_j,u) \end{array}\right) \ , \labbel{eq:udInu} 
\ee 
with 
\begin{align} B_{00}(u) &= 
       \frac{1}{6u} - \frac{1}{u-m^2} + \frac{1}{3(u-9m^2)} \ , \nonumber\\ 
 B_{02}(u) &= \frac{1}{m^4}\left( - \frac{1}{6u} + \frac{3}{16(u-m^2)}
                     - \frac{1}{48(u-9m^2)} \right) \ , \labbel{eq:udI0u} 
\end{align} 
\begin{align} B_{20}(u) &= 
    - \frac{1}{3}m^2 + \frac{1}{6}u + \frac{m^4}{6u}
    - \frac{16m^4}{3(u-m^2)} + \frac{16m^4}{3(m-9m^2)} \ , \nonumber\\ 
              B_{22}(u) &= - \frac{1}{6u} + \frac{1}{u-m^2}
                       -  \frac{1}{3(u-9m^2)} \ , \labbel{eq:udI2u} 
\end{align} 
{\it i.e.} the system decouples into a (rather simple!) equation for the 
function $ I_1(b_i,b_j,u) $ and a two by two first order homogeneous system 
for the two functions $ I_0(b_i,b_j,u), I_2(b_i,b_j,u) $ in which 
$ I_1(b_i,b_j,u) $ does not appear anymore. Let us just recall that 
Eq.(\ref{eq:udI1u}) implies that $ I_1(b_i,b_j,u) $ is constant, 
with the value of the constant depending on the actual choice of 
roots $ (b_i,b_j) $ (see for instance Eq.s(\ref{eq:valI1}) below). 
\par 
The two by two system can be recast in the form of a single 
{\it second order} homogeneous differential equation for one of the two 
functions, say $ I_0(b_i,b_j,u) $; to that aim, we rewrite the first 
of the Eq.s(\ref{eq:udInu}) as 
\be I_2(b_i,b_j,u) = \Duu\ I_0(b_i,b_j,u) \ . \labbel{eq:I2} \ee 
where $ D_1(u,d/du) $ is the first order differential operator 
\be \Duu = \left[ -\frac{2}{3}u(u-m^2)(u-9m^2) \frac{d}{du} 
      + \left(-\frac{1}{3}u^2 + \frac{14}{3}m^2 u + m^4 \right) 
             \right]      \ , \labbel{eq:D1u} 
\ee 
and then evaluate the $u$-derivative of that same first equation of 
(\ref{eq:udInu}). 
By expressing in the result the derivative of $ I_2(b_i,b_j,u) $ through 
the second of the Eq.s(\ref{eq:udInu}) and then $ I_2(b_i,b_j,u) $ through 
Eq.(\ref{eq:I2}), we obtain 
\be \Du \ I_0(b_i,b_j,u) = 0 \ , \labbel{eq:Eq2} \ee 
where $ D(u,d/du) $ is the {\it second order} differential operator 
\begin{align} \Du = \biggl\{ & \frac{d^2}{du^2} + \left[\frac{1}{u} 
            + \frac{1}{u-m^2} + \frac{1}{u-9m^2} \right] 
                                            \frac{d}{du}  \nonumber\\ 
 +&  \frac{1}{m^2}\left[ - \frac{1}{3u} + \frac{1}{4(u-m^2)} 
                   + \frac{1}{12(u-9m^2)} \right] \biggr\} \ . 
\labbel{eq:Du} \end{align} 
\par 
Quite in general, the two by two {\it first order} differential system 
in Eq.(\ref{eq:udInu}) 
\be \frac{d}{du} 
    \left(\begin{array}{c} f_0(u) \\ f_2(u) \end{array}\right) = 
    \left(\begin{array}{cc} 
          B_{00}(u) & B_{02}(u) \\ 
          B_{20}(u) & B_{22}(u) \end{array}\right) 
    \left(\begin{array}{c} f_0(u) \\ f_2(u) \end{array}\right) \ , 
          \labbel{eq:udfnu} 
\ee 
has two pairs of linearly independent solutions, while 
a {\it second order} differential equation like Eq.(\ref{eq:Eq2}) 
\be \Du \ f_0(u) = 0 \labbel{eq:DuI} \ee 
has two linearly independent solutions. 
\par 
As a first solution of Eq.(\ref{eq:DuI}) we can take 
\begin{align}  I_0(u) &= I_0(4m^2,(W-m)^2,u) \nonumber\\ 
           &= \int_{4m^2}^{(W-m)^2} \frac{db}{\sqrt{R_4(u,b)}} \ , 
\labbel{eq:I0ub} \end{align} 
where $ I_0(u) $ is the function already introduced in Eq.(\ref{eq:I0u}). 
For obtaining a second solution of the same equation, let us write 
\be I(0,4m^2,n,u) =  \int_0^{4m^2} \frac{db\;b^n}{\sqrt{-R_4(u,b)}} \ ; 
\labbel{eq:J(n)} \ee 
where we have changed $ \sqrt{R_4(u,b)} $ into $ \sqrt{-R_4(u,b)} $ 
to keep the solution real. 
It is obvious that all the homogeneous relations valid for the generic 
functions $ I(b_i,b_j,n,u) $ apply as well to the functions 
$ I(0,4m^2,n,u) $, as they are equal to the corresponding $ I(b_i,b_j,n,u) $ 
of Eq.s(\ref{eq:I(n,u)}), times an overall imaginary factor $ i $. 
The function $ J_0(u) $, defined as 
\begin{align} J_0(u) &= I(0,4m^2,0,u) \nonumber\\ 
&= \int_0^{4m^2} \frac{db}{\sqrt{-R_4(u,b)}} \ ,
\labbel{eq:J0u} \end{align} 
is therefore a second solution of Eq.(\ref{eq:Eq2}). 
\par 
The $ u \to 9m^2 $ limit of the above function 
(see for instance Eq.s(8.12) of~\cite{Remiddi:2016gno}) is 
\be \lim_{u\to 9m^2+} J_0(u) = \frac{\sqrt{3}}{2m^2} 
    \left(\frac{\ln3}{3} + \frac{\ln2}{2} - \frac{\ln(u-9m^2)}{3} 
    \right) \ , \labbel{eq:J0at9} \ , \ee 
showing in particular, for comparison with Eq.(\ref{eq:I0at9}), that 
the two functions $ I_0(u) $ and $ J_0(u) $ are linearly independent. 
An explicit calculation gives also
\be J_0(u) =\frac{2}{\sqrt{(W+3m)(W-m)^3}} 
           \ K\left(1-\frac{(W-3m)(W+m)^3}{(W+3m)(W-m)^3}\right) \ . 
\labbel{eq:JeqK} \ee 
\par 
Summarizing, we have 
\begin{align} \Du I_0(u) &= 0 \ , \nonumber\\ 
\Du J_0(u) &= 0 \ . \labbel{eq:DuI0J0} \end{align} 
For completeness, we recall (from~\cite{Remiddi:2016gno}) also the values 
of $ I_1(u), J_1(u) $ (which are constant) 
\begin{align} I_1(u) &= I_1(4m^2,(W-m)^2,u) = 0 \ , \nonumber\\ 
           J_1(u) &= I_1(0,4m^2,u) = - \frac{\pi}{3} \ , \labbel{eq:valI1} 
\end{align} 
consistent, of course, with Eq.(\ref{eq:udI1u}). 
\par 
As a further remark, given any solution of the second order equation 
Eq.(\ref{eq:DuI}) corresponding to the two by two system \eqref{eq:udfnu}, 
we can complete the pair of solutions by using Eq.(\ref{eq:D1u}); 
if $ I_0(u), J_0(u) $ are the solutions of the second order 
equation, the accompanying function $ I_2(u), J_2(u) $ are then given by 
\begin{align} I_2(u) &= \Duu\ I_0(u) \ , \nonumber\\ 
              J_2(u) &= \Duu\ J_0(u) \ , \labbel{eq:IJ2u} 
\end{align} 
so that the two independent pairs of solutions of Eq.(\ref{eq:udfnu}) are 
given by the two columns of the matrix 
\be \left(\begin{array}{cc} I_0(u) & J_0(u) \\ 
    I_2(u) & J_2(u) \end{array}\right) \ , 
\labbel{eq:IJ} \ee 
which therefore satisfy (in matricial form) the equations 
\be \frac{d}{du} 
    \left(\begin{array}{cc} I_0(u) & J_0(u) \\ 
       I_2(u) & J_2(u) \end{array}\right) = 
    \left(\begin{array}{cc} 
          B_{00}(u) & B_{02}(u) \\ 
          B_{20}(u) & B_{22}(u) \end{array}\right) 
    \left(\begin{array}{cc} I_0(u) & J_0(u) \\ 
       I_2(u) & J_2(u) \end{array}\right) = 0\,.
\labbel{eq:IJ02} \ee 
For convenience of later use, let us observe here that we have also, 
according to Eq.s(\ref{eq:Inu}) and Eq.s(\ref{eq:valI1}) 
\begin{align} 
   I_2(u) &= \int_{4m^2}^{(W-m)^2} \frac{db\ b^2}{\sqrt{R_4(u,b)}} \ , 
                            \nonumber\\ 
   J_2(u) &= \int_{0}^{4m^2} \frac{db\ b^2}{\sqrt{-R_4(u,b)}} 
           + \frac{\pi}{3}(u+3m^2) \ . 
\labbel{eq:IJ2b} 
\end{align} 
\par 
Let us repeat here that, as anticipated in the Introduction, due to 
Eq.s~\eqref{eq:IJ2u} $ I_k(u) $ and $ J_k(u) $ have E-weight 
equal to zero. \par
In the range $ 9m^2 < u < \infty $ the two functions 
$ I_0(u), J_0(u) $ are real, outside that range they develop also an 
imaginary part and become complex; the details of their analytic 
continuation can be found, although with a slightly different 
notation, in Appendix B of~\cite{Remiddi:2016gno}. 
\par 
We can now look back at the three by three system Eq.(\ref{eq:udI(n)}). 
It has three linearly independent solutions, each solution being a set 
of three functions, namely the three sets $ I(0,4m^2,n,u) $, 
$ I(4m^2,(W-m)^2,n,u) $ and $ I((W-m)^2,(W+m)^2,n,u) $ with 
$ n=0,1,2. $ With the change of basis of Eq.s(\ref{eq:Inu}), 
the two sets $ I(0,4m^2,n,u) $ and $ I(4m^2,(W-m)^2,n,u) $ correspond 
to the decoupled sets $ J_n(u), I_n(u) $ just discussed. \par 
Concerning the third set, 
\be I((W-m)^2,(W+m)^2,n,u) = \int_{(W-m)^2}^{(W+m)^2} db 
                 \frac{b^n}{\sqrt{-R_4(u,b)}} \ , \labbel{eq:K(n)} \ee 
an explicit calculation (based on contour integration arguments in the 
complex plane, see for instance~\cite{Remiddi:2016gno}) gives 
\begin{align} I((W-m)^2,(W+m)^2,0,u) &= I(0,4m^2,0,u) \ , \nonumber\\ 
 I((W-m)^2,(W+m)^2,1,u) &= \frac{u+3m^2}{3}I((W-m)^2,(W+m)^2,0,u) 
                         + \frac{2}{3}\pi \ , \nonumber\\ 
 I((W-m)^2,(W+m)^2,2,u) &= I(0,4m^2,2,u) + (u+3m^2)\pi \ . 
\labbel{eq:Kn} \end{align} 
The transformations  Eq.s(\ref{eq:Inu}) then read 
\begin{align} 
  K_0(u) &= I((W-m)^2,(W+m)^2,0,u) 
       = \int_{(W-m)^2}^{(W+m)^2} \frac{db}{\sqrt{-R_4(u,b)}} \nonumber\\ 
         &= J_0(u) \ , \nonumber\\ 
 K_1(u) &= I((W-m)^2,(W+m)^2,1,u) - \frac{u+3m^2}{3}I((W-m)^2,(W+m)^2,0,u) 
      \nonumber\\ &= J_1(u) + \pi = \frac{2}{3}\pi \ , \nonumber\\ 
 K_2(u) &= I((W-m)^2,(W+m)^2,2,u) - (u+3m^2)I((W-m)^2,(W+m)^2,1,u) 
        \nonumber\\ 
     &+ \frac{(u+3m^2)^2}{3} I((W-m)^2,(W+m)^2,0,u) 
     \nonumber\\ &= J_2(u) \ . 
\labbel{eq:Knu} 
\end{align} 
Note that the pair $ K_0(u), K_2(u) $, being a solution of the two by two 
system (\ref{eq:udfnu}), must be a linear combination of the two already 
discussed pairs of solutions, $ I_0(u), I_2(u) $ and $ J_0(u), J_2(u) $, 
and if fact they are just equal to $ J_0(u) $ and $J_2(u) $, but 
$ K_1(u) $ differs from $ J_1(u) $. 
$I_k(u)$, $J_k(u)$ and $K_k(u)$ for $k=0,1,2$ are therefore indeed
the entries of a $3 \times 3$ matrix of homogeneous solutions
of the $3 \times 3$ system of differential equations
\be
\frac{d}{du} 
    \left(\begin{array}{c} f_0(u) \\ f_2(u) \\ f_1(u) \end{array}\right) = 
    \left(\begin{array}{ccc} 
          B_{00}(u) & B_{02}(u) & 0 \\ 
          B_{20}(u) & B_{22}(u) & 0 \\ 0 & 0 & 0 \end{array}\right) 
    \left(\begin{array}{c} f_0(u) \\ f_2(u) \\ f_1(u) \end{array}\right) \,;
    \labbel{eq:3x3h}
\ee
namely explicitly we have
\be
\frac{d}{du} 
    \left(\begin{array}{ccc} 
          I_0(u) & J_0(u) & J_0(u) \\ 
          I_2(u) & J_2(u) & J_2(u) \\ 0 & -\frac{\pi}{3} & \frac{2\, \pi}{3} \end{array}\right)  = 
    \left(\begin{array}{ccc} 
          B_{00}(u) & B_{02}(u) & 0 \\ 
          B_{20}(u) & B_{22}(u) & 0 \\ 0 & 0 & 0 \end{array}\right) 
    \left(\begin{array}{ccc} 
          I_0(u) & J_0(u) & J_0(u) \\ 
          I_2(u) & J_2(u) & J_2(u) \\ 0 & -\frac{\pi}{3} & \frac{2\, \pi}{3} \end{array}\right) \,,
    \labbel{eq:3x3hexpl}
\ee
where we used  of $K_0(u)=J_0(u)$, $K_2(u)=J_2(u)$, $I_1(u) = 0$, $J_1(u) = -\pi/3$ and
$K_1(u) = 2 \pi /3$.

Besides the homogeneous equations Eq.(\ref{eq:3x3h})
we will consider also the corresponding inhomogeneous equations, namely 
in matrix form
\be \frac{d}{du} 
    \left(\begin{array}{c} g_0(u) \\ g_2(u) \\ g_1(u)\end{array}\right) = 
    \left(\begin{array}{ccc} 
          B_{00}(u) & B_{02}(u) & 0 \\ 
          B_{20}(u) & B_{22}(u) & 0 \\ 0 & 0 & 0 \end{array}\right) 
    \left(\begin{array}{c} g_0(u) \\ g_2(u) \\ g_1(u) \end{array}\right) 
  + \left(\begin{array}{c} N_0(u) \\ N_2(u) \\ N_1(u) \end{array}\right) \ , 
\labbel{eq:udgnu} \ee 
where $ N_0(u), N_1(u), N_2(u) $ are the inhomogeneous terms, supposedly known, 
and the functions $ g_0(u)$, $g_1(u)$, $g_2(u) $ are the unknown. 
The system is equivalent an inhomogeneous second order 
equation for $ g_0(u) $ and a first order differential equation for $g_1(u)$, 
\be \Du g_0(u) = N(u) \ , \labbel{eq:Dug0} \ee 
\be \frac{d}{du} g_1(u) = N_1(u) \labbel{eq:Dug1} \ee
with $ g_2(u) $ given by 
\be g_2(u) = \Duu g_0(u) + N_0(u) \ , \labbel{eq:Dug2} \ee 
where $ D_1(u,d/du) $ is given by Eq.(\ref{eq:D1u}), and $ N(u) $ is related 
to $ N_0(u), N_2(u) $ of Eq.(\ref{eq:udgnu}) by the relation 
\begin{align} N(u) 
  &= \left( \frac{7}{6u} + \frac{4}{3(u-9m^2)} + \frac{d}{du} 
     \right)N_0(u) \nonumber\\ 
  &+ \frac{1}{m^4}\left( - \frac{1}{6u} + \frac{3}{16(u-m^2)} 
                         - \frac{1}{48(u-9m^2)} \right)N_2(u) \ . 
\labbel{eq:N} \end{align} 
\par 
The solution of Eq.s(\ref{eq:udgnu}) can be obtained by 
the Euler-Lagrange method; to that aim, given the decoupled form
of the system, one can split the problem in two steps.
First, one solves the rather trivial first order {\it inhomogeneous} 
differential equation for $g_1(u)$ by quadrature obtaining
\be
g_1(u) = \int^u \, dv N_1(v) + c_1\,,
\ee
where $c_1$ is an integration constant.
Then, in order to solve the two by two coupled system, 
one considers the two by two
matrix of the two independent solutions already introduced in 
Eq.(\ref{eq:IJ}), 
\be \left(\begin{array}{cc} I_0(u) & J_0(u) \\ 
    I_2(u) & J_2(u) \end{array}\right) \ , \labbel{eq:GI0J0} \ee 
and its determinant, the Wronskian of the system, defined as 
\be W_s(u) = I_0(u)J_2(u) - I_2(u)J_0(u) \ . \labbel{eq:Ws} \ee 
From the very definition, it satisfies the equation 
\be \frac{d}{du} W_s(u) = \left( B_{00}(u) + B_{22}(u) \right) W_s(u) 
    = 0 \ , \labbel{eq:udWs} \ee 
where use is made of Eq.s(\ref{eq:udI2u}), showing that $ W_s(u) $ is a 
constant; an explicit calculation gives indeed~\cite{Remiddi:2016gno},  
\be W_s(u) = \pi \ . \labbel{eq:Wspi} \ee 
The inverse of the matrix (\ref{eq:GI0J0}) is therefore 
\be \frac{1}{\pi} \left(\begin{array}{cc} J_2(u) & -I_0(u) \\
    -I_2(u) & I_0(u) \end{array}\right) \ . \labbel{eq:iGI0J0} \ee
The Euler-Lagrange method then gives the solutions of the 
Eq.s(\ref{eq:udgnu}) in the form 
\be \left(\begin{array}{c} g_0(u) \\ g_2(u) \end{array}\right) = 
    \left(\begin{array}{cc} I_0(u) & J_0(u) \\ 
    I_2(u) & J_2(u) \end{array}\right) 
 \biggl[ \left(\begin{array}{c} c_0 \\ c_2 \end{array}\right) 
       + \frac{1}{\pi}\int^u dv 
         \left(\begin{array}{cc} J_2(v) & -J_0(v) \\ 
               -I_2(v) & I_0(v) \end{array}\right) 
         \left(\begin{array}{c} N_0(v) \\ N_2(v) \end{array}\right) 
 \biggr] \ , \labbel{eq:sudgnu} \ee 
where $ c_0, c_2 $ are two more integration constants. 
\par 
Let us note that the above formula can be derived by considering the 
following first order derivatives 
\begin{align} \frac{d}{du}\left[ I_0(u)g_2(u) - I_2(u)g_0(u) \right] 
         &= I_0(u) N_2(u) - I_2(u) N_0(u) \ , \nonumber\\ 
      \frac{d}{du}\left[ J_0(u)g_2(u) - J_2(u)g_0(u) \right] 
         &= J_0(u) N_2(u) - J_2(u) N_0(u) \ , \labbel{eq:psWeq} 
\end{align} 
where $ B_{00}(u) + B_{22}(u) = 0 $ was used. By quadrature one obtains, 
up to constants, 
\begin{align} I_0(u)g_2(u) - I_2(u)g_0(u) &= 
       \int^u dv \left[ I_0(v) N_2(v) - I_2(v) N_0(v) \right] +c_0\ , \nonumber\\ 
 J_0(u)g_2(u) - J_2(u)g_0(u) &= 
       \int^u dv \left[ J_0(v) N_2(v) - J_2(v) N_0(v) \right] + c_2\ , 
\labbel{eq:psWsol} \end{align} 
from which Eq.s\eqref{eq:sudgnu} are immediately recovered. Such a procedure, 
while of course fully equivalent to the Euler-Lagrange method, may provide 
with a different way of grouping terms in the intermediate results. \par 
In a similar way, one can also solve the second order equation 
Eq.(\ref{eq:Dug0}). As a first step, one introduces the Wronskian $ W(u) $ 
of the two solutions of the equation, defined as 
\be W(u) = I_0(u)\frac{d}{du}J_0(u) - J_0(u)\frac{d}{du}I_0(u) \ , 
\labbel{eq:Wr} \ee 
which satisfies the first order homogeneous equation 
\be \frac{d}{du} W(u) = - \left[\frac{1}{u} + \frac{1}{u-m^2} 
       + \frac{1}{u-9m^2} \right] W(u) \ . \labbel{eq:eqWr} \ee 
Its solution, with $ I_0(u), J_0(u) $ given by 
Eq.s(\ref{eq:I0u},\ref{eq:J0u}), is 
\be W(u) = - \frac{3\pi}{2u(u-m^2)(u-9m^2)} \ . \labbel{eq:VWr} \ee 
The solution of Eq.(\ref{eq:Dug0}) then reads 
\begin{align} g_0(u) &= 
   I_0(u) \left( c_0 - \int^u \frac{dv}{W(v)}J_0(v)N(v) \right) 
 + J_0(u) \left( c_2 + \int^u \frac{dv}{W(v)}I_0(v)N(v) \right) 
                                                           \nonumber\\ 
 &= c_0 I_0(u) + c_2 J_0(u) - \int^u \frac{dv}{W(v)} 
           \biggl[ I_0(u) J_0(v) - J_0(u)I_0(v) \biggr] N(v) \ , 
                                                           \labbel{eq:duIJN} 
\end{align} 
where the two integration constants $ c_0, c_2$, to be fixed by the 
boundary conditions, are the same as in Eq.(\ref{eq:sudgnu}); 
$ g_2(u) $ is then given by Eq.(\ref{eq:Dug2}). 
\section{Repeated integrations of $ I_0(u), J_0(u) $ and rational factors } 
\labbel{sec:E1pol} \setcounter{equation}{0} 
\numberwithin{equation}{section} 
In the previous Section, we have introduced the pairs of functions 
$ I_0(u), I_2(u) $ and $ J_0(u), J_2(u) $, and shown their use in 
writing the solution Eqs.(\ref{eq:sudgnu},\ref{eq:duIJN}) of the 
inhomogeneous equations Eqs.(\ref{eq:udgnu},\ref{eq:Dug0}). 
As it is easy to imagine, the integration of products of those 
functions times rational factors appears even in the simplest cases.
Therefore, before studying the more general E-polylogarithms,
we consider now the properties of such (possibly repeated) integrations, 
discussing the analogy with the ordinary generalized 
polylogarithms~\cite{Goncharov,Remiddi:1999ew,Gehrmann:2000zt,
Vollinga:2004sn}, also called Goncharov functions, 
of common use in the evaluation of Feynman graph amplitudes. 
The Goncharov functions can be defined as 
\be G^{[n]}(p_n,p_{n-1},..,p_1;u) = \int_{u_0}^u \frac{du_n}{u_n-p_n} 
  \int_{u_0}^{u_n} \frac{du_{n-1}}{u_{n-1}-p_{n-1}} ..... 
  \int_{u_0}^{u_2} \frac{du_{1}}{u_1-p_{1}} \ , \labbel{eq:Gnu} \ee 
where the parameters $ p_i $ vary within a given finite set of values, 
proper of the problem under study. The repeated integrations arise 
naturally when solving iteratively the differential equations by the Euler 
approach ({\it i.e.} evaluating first the solution of the homogeneous equation 
and then accounting of the inhomogeneous term with the variation of the 
constants method). 
The superscript $ n $ is called the degree (or polylogarithmic weight) 
of the function. In the context of this paper, we will refer to this weight as
$G$-weight for obvious reasons.
\par 
By construction, these functions satisfy the relation 
\be \frac{d}{du} G^{[n]}(p_n,p_{n-1},..,p_1;u) = 
                 \frac{1}{u-p_n}\ G^{[n-1]}(p_{n-1},..,p_1;u) \ , 
\labbel{eq:udGnu} \ee 
{\it i.e.} the derivative of a function of $G$-weight $ n $ is a function of 
the same family but of lower weight $ n-1 $ (times a rational factor). 
For completeness, we can define also 
\be G^{[0]}(u) = 1 \ , \labbel{eq:G0u} \ee 
which satisfies, obviously, the equation 
\be \frac{d}{du} G^{[0]}(u) = 0 \ , \labbel{eq:udG0u} \ee 
such that a function of $G$-weight equal to zero is annihilated
by the first order differential operator $d/du$.

By following as much as possible Eq.(\ref{eq:Gnu}), we start considering the 
functions defined by repeated integrations for integer $ n>0 $ as follows
\begin{align}  
   I_k^{[n]}(p_n,p_{n-1},..,p_1;u) &= \int_{u_0}^u \frac{du_n}{u_n-p_n} 
  \int_{u_0}^{u_n} \frac{du_{n-1}}{u_{n-1}-p_{n-1}} ..... 
  \int_{u_0}^{u_2} \frac{du_{1}}{u_1-p_{1}} I_k(u_1) \ , \nonumber\\ 
   J_k^{[n]}(p_n,p_{n-1},..,p_1;u) &= \int_{u_0}^u \frac{du_n}{u_n-p_n} 
  \int_{u_0}^{u_n} \frac{du_{n-1}}{u_{n-1}-p_{n-1}} ..... 
  \int_{u_0}^{u_2} \frac{du_{1}}{u_1-p_{1}} J_k(u_1) \ , \labbel{eq:IJnu} 
\end{align} 
where the index $ k $ takes the two values $ k= 0 $ and $ k=2 $, 
with the $ p_i $ taking any of the values of the set $\{0,m^2,9m^2\}$. 
Clearly, for $n > 0$ these functions behave very similarly to the
$G$-functions under differentiation, such that one would be tempted
to associate to them a $G$-weight in the same way.
Nevertheless, as already noted, for $ n= 0 $, one defines $ G^{[0]}(u) = 1 $ 
so that  $ d/du(G^{[0]}(u)) = 0 $, Eq.s(\ref{eq:G0u},\ref{eq:udG0u}); 
on the contrary, 
the definition of $ I_k^{[n]}(..,u) $ cannot be na\"ively extended to 
$ n = 0 $, defining for instance $ I_k^{[0]}(u) $ equal to $ I_k(u) $, 
because, at variance with Eq.(\ref{eq:G0u}), 
\be \frac{d}{du} I_k(u) \ne 0 \ , \labbel{eq:udIJ0u} \ee 
and therefore $I_k^{[0]}(u)$ or $J_k^{[0]}(u)$ do not 
have zero $G$-weight. For this reason, without any claim of rigour
or completeness, we call the weight of these functions $E_0$-weight,
in order to clearly distinguish it from the standard polylogarithmic G-weight,
but also from the more general E-weight of E-polylogarithms.
\par
The first of Eq.s(\ref{eq:IJnu}) might also be written, recursively, as 
\be I_k^{[n]}(p_n,p_{n-1},...,p_1;u) = \int_{u_0}^u \frac{du_n}{u_n-p_n} 
    I_k^{[n-1]}(p_{n-1},...,p_1;u_n) \ , \labbel{eq:E1Inu1} \ee 
from which one has at once, for $ n > 1 $,  
\be \frac{d}{du} I_k^{[n]}(p_n,p_{n-1},...,p_1;u) =  \frac{1}{u-p_n} 
    I_k^{[n-1]}(p_{n-1},...,p_1;u) \ , \labbel{eq:udE1Inu} \ee 
which is the straightforward equivalent of Eq.(\ref{eq:udGnu}); but the above 
equation is valid only for $ n> 1 $, because $ I_k^{[n]}(..,u) $ is not 
defined for $ n= 0 $. 
(The same equations hold, obviously, for $ J_k^{[n]}(...;u) $ as well.) 
\par 
Note that the rational factors appearing in the previous definitions 
are all of the form $ 1/(u-p) = d/du\{\ln(u-p)\} $; different 
powers of those factors, such as for instance $ 1, v, 1/(v-p)^2 $ 
{\it etc.} can be integrated by parts, without increasing the 
weight of the function, as for instance in the following example 
(valid for $ n>1 $) 
\begin{align} 
   \int^u dv\ I_k^{[n]}(p_n,...,p_1;v) &= \int^u dv \int_{u_0}^v 
           \frac{dv_n}{v_n-p_n} I_k^{[n-1]}(p_{n-1},...,p_1;v_n) \nonumber\\ 
  &= \int_{u_0}^v dv \biggl( \frac{u-p_n}{v-p_n} - 1 \biggr) 
    I_k^{[n-1]}(p_{n-1},...,p_1;v) \nonumber \ , 
\end{align} 
and the procedure can be used recursively, down to $ n=2 $. \par 
For $ n = 1 $, however, the integrations by parts involve the $ u $ 
derivatives of $ I_k(u) $, which are non zero and can instead
be expressed in terms of the same functions times a combination of the same 
rational factors, see Eq.s(\ref{eq:udInu}). The direct, na\"ive 
integration-by-parts approach is therefore not sufficient in the case 
of the very first integration involving $ I_k(v) $ or $ J_k(v) $; 
indeed, one has rather to write the 
complete {\it system} of integration by parts identities obtained by 
considering the products of all the powers of the 
rational factors times the functions $ I_k(v) $ or $ J_k(v) $, and then to 
solve the system in terms of the {\it master integrals} of the problem. 
The generic identity has the (obvious) form 
$$ \int^u dv \frac{d}{dv}X(v) = X(u) - X(u_0) \ , $$ 
where $ X(v) $ stands for the products of all the possible factors 
$\{ 1, v^n, 1/v^n , 1/(v-m^2)^n , 1/(v-9m^2)^n \}$ times $ I_k(u) $ or 
$ J_k(u) $, and $ n $ is any positive integer. 
\par 
As a result, it turns out that {\it all} the integral of the form 
$$ \int^u dv \ X(v) \ , $$ 
where $ X(v) $ was just defined above, including both $ I_k(v) $ and 
$ J_k(v) $, can be expressed in terms of the four {\it master integrals} 
\be \int^u dv \biggl( 1 \ ; \frac{1}{v} \ ; \frac{1}{v-m^2} \ ; 
                  \frac{1}{v-9m^2} \ \biggr)\ I_0(v) \ , \labbel{eq:E1masters}
                  \ee
which involve only $ I_0(v) $, plus terms in $ I_k(u) $ (not integrated) 
generated by the integration by parts. \par 
A few examples (written as relations among indefinite integrals, 
{\it i.e.} valid up to a constant) are 
\begin{align} 
  \int^u dv\ I_2(v) &= 6m^4 \int^u dv\ I_0(v) \nonumber\\ 
     &+ \biggl( \frac{1}{24}m^2u^2 - \frac{223}{24}m^4u - \frac{11}{8}m^6 
               - \frac{1}{24}u^3 \biggr) I_0(u) 
      + \biggl( \frac{11}{8}m^2 + \frac{5}{8}u \biggr) I_2(u) 
                         \ , \nonumber\\ 
  \int^u dv\ v\ I_0(v) &= 3m^2\int^u dv\ I_0(v) \nonumber\\ 
     &+ \biggl( -7m^2u - \frac{3}{2}m^4 + \frac{1}{2} u^2 \biggr)I_0(u) 
      + \frac{3}{2} I_2(u) \ , \nonumber\\  
  \int^u dv\ v^2\ I_0(v) &= 15m^4\int^u dv\ I_0(v)  \nonumber\\ 
   &+ \biggl( \frac{11}{8}m^2 u^2 - \frac{307}{8} m^4 u - \frac{69}{8}m^6 
    + \frac{3}{8} u^3 \biggr) I_0(u) 
    + \biggl( \frac{69}{9}m^2 + \frac{3}{8} u \biggr) I_2(u) 
                         \ , \nonumber\\ 
   \int^u dv\ \frac{1}{v-m^2}\ I_2(v) &= 
   \int^u dv\ \biggl( \frac{4}{3}m^2 + \frac{16}{3}\ \frac{m^4}{v-m^2} 
              \biggr) I_0(v) \nonumber\\ 
   &+ \biggl( -m^2u - \frac{3}{2}m^4 - \frac{1}{6} u^2 \biggr) I_0(u) 
    + \frac{3}{2} I_2(u) \ . \labbel{eq:E1ibp} 
\end{align} 
\par 
For the analytical expression of the master integrals of 
Eq.\eqref{eq:E1masters} we refer to Appendix~\ref{App:Masters}. 
So far we have considered repeated integrations associated to the pair of 
functions $ I_0(u), I_2(u) $; obviously, the procedure applies as well to 
the other pair of functions, $ J_0(u), J_2(u) $, which satisfies the 
same homogeneous equations as $ I_0(u), I_2(u) $. While the equations 
(\ref{eq:E1ibp}) (defined up to a constant) remain valid under the exchange 
of the two pairs of functions, the explicit expression of the four 
{\it master integrals}, corresponding to Eq.s 
(\ref{eq:lnS},\ref{eq:lnT},\ref{eq:lnU},\ref{eq:lnV}) is of course 
different. \par 
As a final remark for this Section, consider a function 
$ GI_n(u) $ of the form 
\be GI_0^{[n]}(u) = I_0(u) \ G^{[n]}(u) \ , \labbel{eq:GIn} \ee 
where $ G^{[n]}(u) $ is a function of either $G$- or $E_0$-weight 
$ n $ in the sense defined above,
(i.e. obtained by $n$ repeated integrations over rational functions) 
and $ I_0(u) $ is once more the function of 
Eq.(\ref{eq:I0u})\footnote{The discussion applies, of course, to $ J_0(u) $ 
Eq.(\ref{eq:J0u}) as well, but not to $ I_2(u) $ or $ J_2(u) $.}; 
let us further recall that $ I_0(u), J_0(u) $ do not possess definite 
$G$- or $E_0$-weight, 
so that $ G^{[n]}(u) $ cannot be $ I_0(u) $ or a product of $ I_0(u) $ and 
$ J_0(u) $. 
Recalling Eq.s(\ref{eq:DuI0J0}) an elementary calculation gives
\be \Du GI_0^{[n]}(u) = \sum_{k=0,2}r^{[n-1]}_{0,k}(u) GI_k^{[n-1]}(u) + 
\sum_{k=0,2}r^{[n-2]}_{0,k}(u)GI_k^{[n-2]}(u) \ , 
\labbel{eq:DuGIn} \ee 
where the $r^{[j]}_{0,k}(u)$ are simple rational functions.
The above equation shows that a function of the form (\ref{eq:GIn}), 
with $ G^{[n]}(u) $ of $G$- or $E_0$-weight $ n $, satisfies Eq.(\ref{eq:EpolDudef}) 
with an inhomogeneous term which contains only first and second derivatives of 
$ G^{[n]}(u) $, and therefore contains terms of weight $n-1$ and $n-2$.
This implies that these functions $ GI^{[n]}(u)$ are indeed E-polylogarithms
with E-weight equal to $n$.
This observation will be useful in the next sections.


\section{A first example of an E-polylogarithm} 
\labbel{sec:FF} \setcounter{equation}{0} 
\numberwithin{equation}{section} 
Having discussed in detail the properties of the functions $I_0(u)$, $J_0(u)$ 
and of (na\"ive) iterative integrations over the latter with rational 
functions, we are now ready to consider the main topic 
of this paper. Let us start with an explicit example, namely the function
\be \EI_0^{[1]}(0,u) =  \int_{4m^2}^{(W-m)^2} \frac{db}{\sqrt{R_4(u,b)}} 
            \ G(0,b)  =  \int_{4m^2}^{(W-m)^2} \frac{db}{\sqrt{R_4(u,b)}} 
            \ \ln{b} \labbel{eq:Gl0u} \ee 
(where, for simplicity, instead of $ \ln(b/m^2) $ we have written $ \ln{b} $), 
whose value at $ u=9m^2 $ is 
\be \lim_{u\to 9m^2} \EI_0^{[1]}(0,u) = \frac{\sqrt{3}}{6m^2}\pi\ln2 \ , 
\labbel{eq:Gl0at9} \ee 
as can be easily checked by using the change of variable (\ref{eq:btot}). 
Let us shortly comment the somewhat clumsy notation used; in the name 
$ \EI_0^{[1]}(0,u) $, $ \EI $ stands for Elliptic integral corresponding 
to the integration range $ 4m^2 < b < (W-m)^2 $, associated to the 
functions $I_k(u)$, the superscript 
$ [1] $ refers to the weight of the (poly)logarithm $ G(0,b)=\ln{b} $, the 
arguments $ (0,u) $ refer to the ``letter" $ 0 $ of the (poly)logarithm and 
(obviously) to the variable $ u $, finally the lower index $ 0 $ is the analog 
of the index $ k = 0 $ of $ I_0(u) $ in Eq.(\ref{eq:I0u}). In 
this notation, one would have 
\be \EI_k^{[0]}(u) = I_k(u) \ , \hspace{2cm} \EJ_k^{[0]}(u) = J_k(u) \ . 
\labbel{eq:EIJk0} \ee 

We will work out this example in detail and outline how this generalizes 
then to higher weights. 
We can start by deriving a second order differential equation for 
$ \EI_0^{[1]}(0,u) $, by following closely the derivation discussed in 
Section \ref{sec:Thebeg}. In analogy with Eq.(\ref{eq:I(n,u)}) we introduce 
the auxiliary functions 
\be Il(b_i,b_j,n,u) = \int_{b_i}^{b_j} 
    \frac{db\ b^n}{\sqrt{R_4(u,b)}} \ \ln{b} \ , 
\labbel{eq:Gl(n,u)} \ee 
such that clearly
$$\EI_0^{[1]}(0,u) = Il(4m^2,(W-m)^2,0,u) = I_0(u)\,.$$
In analogy with Eq.(\ref{eq:dbbd}) one has the identities 
\be \int_{b_i}^{b_j} db \frac{d}{db}\left(\sqrt{R_4(u,b)}\ b^n 
         \ln{b} \right) = 0 \ . \labbel{eq:dbbdGl} \ee 
When the $b$-derivative acts on $ (\sqrt{R_4(u,b)}\ b^n) $, as in the case of 
Eq.(\ref{eq:dbbd}), it generates the same terms, now multiplied also by 
$ \ln{b} $, while, when it acts on the logarithm, it replaces the 
logarithm by the factor $ 1/b $. 
When writing afterwards $ \sqrt{R_4(u,b)} = R_4(u,b)/ \sqrt{R_4(u,b)} $, 
the factor $ b $ present in $ R_4(u,b) $, Eq.(\ref{eq:R4}), cancels 
against the factor $ 1/b $ from the derivative of the logarithm, 
generating the same quantities $ I(b_i,b_j,n,u) $, already introduced in 
Eq.(\ref{eq:I(n,u)}). \par 
Working out the algebra, we are left with an equation, corresponding to 
Eq.(\ref{eq:idbn}), whose {\it l.h.s.} is the {\it l.h.s.} of 
Eq.(\ref{eq:idbn}) with the functions $ I(b_i,b_j,n,u) $ replaced by 
$ Il(b_i,b_j,n,u) $, while the {\it r.h.s.} is no longer vanishing, but 
contains a combination of the $ I(b_i,b_j,n,u) $, due the $b$-derivative 
of $\ \ln{b} $ in Eq.(\ref{eq:dbbdGl}). 
That equation can be used for expressing any 
$ Il(b_i,b_j,n,u) $, with $n $ integer and $ n > 2 $, in terms of the 
three {\it master integrals} $ Il(b_i,b_j,k,u) $, $ k=0,1,2 $; 
the homogeneous part of the relations, {\it i.e.} the part containing 
the $ Il(b_i,b_j,n,u) $, has the same coefficients appearing in 
Eq.(\ref{eq:idbn}), but in the case of the $ Il(b_i,b_j,n,u) $ 
there are also inhomogeneous terms, {\it i.e.} terms containing not 
the $ Il(b_i,b_j,n,u) $ but the $ I(b_i,b_j,n,u) $. 
\par 
We can continue by introducing, in analogy with Eq.(\ref{eq:Qku}), 
the auxiliary quantities 
$$ Ql(k,u) =  \int_{b_i}^{b_j} db \left( \sqrt{R_4(u,b)}\ b^k \right) 
                               \ln{b} \ , $$ 
differentiating them with respect to $ u $ {\it etc.}, we arrive, in analogy 
to Eq.(\ref{eq:udI(n)}), to the following three by three linear system 
of first order differential equations: 
\begin{align}  
    \frac{d}{du} \left(\begin{array}{c} 
              Il(b_i,b_j,0,u) \\ 
              Il(b_i,b_j,1,u) \\ 
              Il(b_i,b_j,2,u) \end{array}\right) &= 
    \left(\begin{array}{ccc} 
          C_{00}(u) & C_{01}(u) & C_{02}(u) \\ 
          C_{10}(u) & C_{11}(u) & C_{12}(u) \\ 
          C_{20}(u) & C_{21}(u) & C_{22}(u) \end{array}\right) 
     \left(\begin{array}{c} 
              Il(b_i,b_j,0,u) \\ 
              Il(b_i,b_j,1,u) \\ 
              Il(b_i,b_j,2,u) \end{array}\right) \nonumber\\ 
  &+ \left(\begin{array}{ccc} 
          C_{00}^{(1)}(u) & C_{01}^{(1)}(u) & C_{02}^{(1)}(u) \\ 
          C_{10}^{(1)}(u) & C_{11}^{(1)}(u) & C_{12}^{(1)}(u) \\ 
          C_{20}^{(1)}(u) & C_{21}^{(1)}(u) & C_{22}^{(1)}(u) 
          \end{array}\right) 
\left(\begin{array}{c} 
               I(b_i,b_j,0,u) \\ 
               I(b_i,b_j,1,u) \\ 
               I(b_i,b_j,2,u) \end{array}\right) \ , \labbel{eq:udGl(n)} 
\end{align} 
where the coeffcients of the homogeneous part, the $ C_{nk}(u) $ are 
the same as in Eq.(\ref{eq:udI(n)}), while the $ C_{nk}^{(1)}(u) $ 
are new, similar coefficients (which we do not write here for brevity) 
multiplying the $ I(b_i,b_j,n,u) $. \par 
Following Eq.s(\ref{eq:Inu}), we introduce a new basis of {\it master 
integrals} with the definitions 
\begin{align} 
Il_0(b_i,b_j,u) &= Il(b_i,b_j,0,u) \nonumber\\ 
     &= \int_{b_i}^{b_j}\frac{db}{\sqrt{R_4(u,b)}} \ln{b} \ , \nonumber\\ 
Il_1(b_i,b_j,u) &= Il(b_i,b_j,1,u) - \frac{u+3m^2}{3} Il(b_i,b_j,0,u) 
                                                              \nonumber\\ 
     &= \int_{b_i}^{b_j}\frac{db}{\sqrt{R_4(u,b)}} 
        \left(b - \frac{u+3m^2}{3} \right)         \ln{b} \ , \nonumber\\ 
Il_2(b_i,b_j,u) &= Il(b_i,b_j,2,u) - (u+3m^2) Il(b_i,b_j,1,u) 
                        + \frac{(u+3m^2)^2}{3} Il(b_i,b_j,0,u) \nonumber\\ 
     &= \int_{b_i}^{b_j}\frac{db}{\sqrt{R_4(u,b)}} 
        \left( b^2 - (u+3m^2)b + \frac{(u+3m^2)^2}{3} \right) 
                                                   \ln{b} \ . \labbel{eq:Glnu} 
\end{align} 
In terms of the functions of the new basis and of the $ I_n(b_i,b_j,u) $ 
the system becomes 
\begin{align} 
\frac{d}{du}Il_1(b_i,b_j,u) &= 
   \left( \frac{2}{9} - \frac{16m^2}{9(u-m^2)} \right) I_0(b_i,b_j,u) 
                                                            \nonumber\\ 
            &+ \frac{2}{3(u-m^2)} I_1(b_i,b_j,u) \   \labbel{eq:udGl1u} 
\end{align} 
and 
\begin{align} 
    \frac{d}{du} \left(\begin{array}{c}
              Il_0(b_i,b_j,u) \\
              Il_2(b_i,b_j,u) \end{array}\right) &=
    \left(\begin{array}{cc}
          B_{00}(u) & B_{02}(u) \\
          B_{20}(u) & B_{22}(u) \end{array}\right)
\left(\begin{array}{c}
              Il_0(b_i,b_j,u) \\
              Il_2(b_i,b_j,u) \end{array}\right) \nonumber\\ 
  &+ \left(\begin{array}{cc}
          B_{00}^{(1)}(u) & B_{02}^{(1)}(u) \\
          B_{20}^{(1)}(u) & B_{22}^{(1)}(u) \end{array}\right)
\left(\begin{array}{c}
              I_0(b_i,b_j,u) \\
              I_2(b_i,b_j,u) \end{array}\right) \ . \labbel{eq:udGlnu} 
\end{align} 
A few comments are in order. Concerning Eq.(\ref{eq:udGl1u}), 
it is to be noted that none of the functions $ Il_k(b_i,b_j,u) $ appears 
in the {\it r.h.s.}, so that the equation, even if not as simple as 
Eq.(\ref{eq:udI1u}), is anyhow a rather trivial differential equation. 
\par 
Concerning Eq.(\ref{eq:udGlnu}), the coefficients of the homogeneous 
part, $ B_{nk}(u) $, are the same as in Eq.(\ref{eq:udInu}), while 
the $ B_{nk}^{(1)}(u) $ (which have a similar structure and are not written 
here again for brevity) are the coefficients of the inhomogeneous terms 
containing the $ I_n(b_i,b_j,u) $. 
Again, the two by two system can be recast in the form of a single 
{\it second order} homogeneous differential equation for 
$ Il_0(b_i,b_j,u) $; the result can be written as 
\begin{align} \Du \ Il_0(b_i,b_j,u) &= \frac{1}{m^2}\biggl( 
  - \frac{8}{9u} + \frac{3}{4(u-m^2)} + \frac{5}{36(u-9m^2)} 
  - \frac{4m^2}{3(u-m^2)^2} \biggr) I_0(b_i,b_j,u) \nonumber\\ 
 &+ \frac{1}{m^6}\biggl( \frac{2}{9u} - \frac{7}{32(u-m^2)} 
  - \frac{1}{288(u-9m^2)} + \frac{4m^2}{(u-m^2)^2} 
    \biggr) I_2(b_i,b_j,u) \ ,  \labbel{eq:EqGl0} 
\end{align} 
while $ Il_2(b_i,b_j,u) $ is given by 
\begin{align} Il_2(b_i,b_j,u) &= \Duu \ Il_0(b_i,b_j,u) \nonumber\\ 
    &+ \left( - \frac{10}{3}m^2u + m^4 + \frac{5}{9}u^2 \right) 
                                         I_0(b_i,b_j,u) \nonumber\\ 
    &+\frac{2}{3} \left( 3m^2 + u \right) I_1(b_i,b_j,u) \nonumber\\ 
    &- I_2(b_i,b_j,u) \ . \labbel{eq:Gl2} 
\end{align} 
The differential operators $ D(u,d/du), D_1(u,d/du) $ in the two above 
equations are of course the same as those defined in 
Eq.s(\ref{eq:Du},\ref{eq:D1u}). 
\par 
We can now specialize the formulas to the case $ b_1 = 4m^2, b_2 =(W-m)^2 $ 
with $ W^2 = u > 9m^2 $, {\it i.e.}, in the notation of Eq.(\ref{eq:Gl0u}), 
$$ \EI_0^{[1]}(0,u) = Il_0(4m^2,(W-m)^2,u) =  \int_{4m^2}^{(W-m)^2} 
     \frac{db}{\sqrt{R_4(u,b)}} \ \ln{b} \ . $$ 
By recalling also Eq.s(\ref{eq:I0ub}) and (\ref{eq:IJ2u}), 
we find finally that Eq.(\ref{eq:EqGl0}) becomes
\begin{align} \Du \ \EI_0^{[1]}(0,u) &= \frac{1}{m^2}\biggl( 
  - \frac{8}{9u} + \frac{3}{4(u-m^2)} + \frac{5}{36(u-9m^2)} 
  - \frac{4m^2}{3(u-m^2)^2} \biggr) I_0(u) \nonumber\\ 
 &+ \frac{1}{m^6}\biggl( \frac{2}{9u} - \frac{7}{32(u-m^2)} 
  - \frac{1}{288(u-9m^2)} + \frac{4m^2}{(u-m^2)^2} 
    \biggr) I_2(u) \ , \labbel{eq:EqGl0b} 
\end{align} 
where, according to the definitions Eq.(\ref{eq:EIJk0}), we can write in the 
{\it r.h.s.}, instead of $ I_k(u) $, $ \EI_k^{[0]}(u) = I_k(u) $. 
As the homogeneous solutions of this equation are known, one can solve 
Eq.(\ref{eq:EqGl0b}) with the help of Eq.(\ref{eq:duIJN}).
The solution clearly reads
\begin{align}
\EI_0^{[1]}(0,u) = c_1^{(1)} I_0(u) + c_2^{(1)} J_0(u)
- \int^u \frac{dv}{W(v)}\Big[ I_0(u) J_0(v) - J_0(u) I_0(v) \Big] 
 N_0(1;u) \,,
\end{align}
with
\begin{align}
N_0(1;u) &=\frac{1}{m^2}\biggl( 
  - \frac{8}{9u} + \frac{3}{4(u-m^2)} + \frac{5}{36(u-9m^2)} 
  - \frac{4m^2}{3(u-m^2)^2} \biggr) I_0(u) \nonumber\\ 
 &+ \frac{1}{m^6}\biggl( \frac{2}{9u} - \frac{7}{32(u-m^2)} 
  - \frac{1}{288(u-9m^2)} + \frac{4m^2}{(u-m^2)^2} 
    \biggr) I_2(u) 
\end{align}
and $W(u)$ is the wronskian given in Eq.~\eqref{eq:VWr}, 
whose value we remind here
$$W(u) = -\frac{3 \pi}{2 \, u \,(u-m^2)(u-9m^2)}\,.$$

Substituting explicitly the value of the Wronskian and 
the result at weight zero we are left with \\
\begin{align}
\EI_0^{[1]}(0,u)  = c_1^{(1)} I_0(u) + c_2^{(1)} J_0(u)
&+ \frac{4}{9\, \pi}\,  \int^u dv\, F_{0,0}(u,v)\, \left[ 
\frac{v}{m^2} + 4 +  \frac{16\, m^2}{(v-m^2)} \right] I_0(v) \nonumber \\
&- \frac{4}{3\, \pi} \int^u dv\, F_{0,0}(u,v)\, 
 \frac{1}{(v-m^2)}I_2(v)\,, \label{eq:resultone}
\end{align}
where we introduced the compact notation
\begin{align}
F_{0,0}(u,v) &=  I_0(u) J_0(v) - J_0(u) I_0(v)   \,.
\end{align}
We need therefore to understand integrals of the form
\begin{equation}\int^u\, dv 
\left\{ 1\;;\; v^n \;;\; \frac{1}{v^n} \;;\; \frac{1}{(v-m^2)^n} \,;\, 
\frac{1}{(v-9m^2)^n} \right\}
F_{0,0}(u,v)\, \Big\{ I_0(v) \, ; \, I_2(v) \Big\}\,.
\end{equation}
Not all these integrals are linearly independent, as we will we show now
by using integration by parts identities. In order to see this, 
let us define the other function
\begin{align}
F_{0,2}(u,v) &=  I_0(u) J_2(v) - J_0(u) I_2(v)  \,,
\end{align}
such that, in the notation of \eqref{eq:udInu}, 
\begin{align}
&\frac{d}{dv} F_{0,0}(u,v) = B_{00}(v) F_{0,0}(u,v) + B_{02}(v) F_{0,2}(u,v) 
\nonumber \\
&\frac{d}{dv} F_{0,2}(u,v) = B_{20}(v) F_{0,0}(u,v) + B_{22}(v) 
F_{0,2}(u,v)\,.
\end{align}
By using  Eq.~\eqref{eq:Wspi}
it is easy to see that
\begin{align}
F_{0,2}(u,v) I_0(v) &= \pi\, I_0(u) + I_2(v) F_{0,0}(u,v) \labbel{eq:one}
\end{align}
such that,
by choosing to re-express $F_{0,2}(u,v) I_0(v)$ in terms of $I_2(v) F_{0,0}(u,v)$, 
we see we should generate all integration by parts identities of the form

\begin{equation*}
\int^u\, dv \, \frac{d}{dv}\; \left( \left\{
1;\; v^n \;; \; \frac{1}{v^n} \;;\; \frac{1}{(v-m^2)^n} \;;\; 
\frac{1}{(v-9m^2)^n} \right\} F_{0,0}(u,v)\,I_0(v) \right) = X_1(u)\,,
\end{equation*}
\begin{equation*}
\int^u\, dv \, \frac{d}{dv}\; \left( \left\{
1;\; v^n \;; \; \frac{1}{v^n} \;;\; \frac{1}{(v-m^2)^n} \;;\; 
\frac{1}{(v-9m^2)^n} \right\} F_{0,0}(u,v)\,I_2(v)  \right) = X_2(u)\,,
\end{equation*}
\begin{equation*}
\int^u\, dv \, \frac{d}{dv}\; \left( \left\{
1;\; v^n \;; \; \frac{1}{v^n} \;;\; \frac{1}{(v-m^2)^n} \;;\; 
\frac{1}{(v-9m^2)^n} \right\} F_{0,2}(u,v)\,I_2(v)\right) = X_3(u)\,,
\end{equation*}
where the $X_j(u)$ are appropriate boundary terms; note that, for simplicity,
we write the IBPs as relations among primitives, i.e. without specifying the lower
integration boundary. This means that all relations we provide here are 
given up to boundary terms. By proceeding 
similarly to the general algorithm described in~\cite{Laporta:2001dd},
we generate a large number of identities for different numerical values 
of the powers $n$ and solve the system of equations. 
We find in this way that all integrals can be expressed in terms of 6 
{\it master integrals}, which we choose as follows
\begin{align}
\int^u\, dv \,   \left\{ 1\,; \,v\,; \,v^2\,;\, \frac{1}{v}\, ;\, 
\frac{1}{v-m^2} \,;\, \frac{1}{v-9 m^2}  \right\} \,
F_{0,0}(u,v)\,I_0(v)\, \labbel{eq:Mastersw1}
\end{align}
plus simpler terms, i.e. terms which do not require integrating over the 
functions $F_{0,0}(u,v)$ and $F_{0,2}(u,v)$.
In particular, we find that one of the integrals in Eq.~\eqref{eq:resultone} 
can be re-expressed as linear combination of the other three as follows

\begin{align}
\int^u\, \frac{dv}{v-m^2} \, F_{0,0}(u,v)\,I_2(v) 
&= \frac{1}{3} \int^u\, dv \, \left( 4 m^2 +  \,v\, 
+ \frac{16 m^4}{v-m^2} \right) F_{0,0}(u,v)\,I_0(v) \nonumber \\
&- \frac{\pi}{2} I_0(u)\,  \int^u \frac{dv}{v-m^2} 
\end{align}
where we see the appearance of a simpler integral, which reminds of the 
shuffle identities for polylogarithms. We stress again, that these relations are
given up to boundary terms. 
By using this identity in 
Eq.~\eqref{eq:resultone} we find at once
\begin{align}
\EI_0^{[1]}(0,u) &= \int_{4m^2}^{(W-m)^2} \kern-15pt 
  \frac{db}{\sqrt{R_4(u,b)}} \ \ln{b}  = c_1^{(1)} I_0(u) + c_2^{(1)} J_0(u)
+ \frac{2}{3}\, \ln{(u-m^2)} I_0(u) \nonumber \\
&= \frac{2}{3}\, \ln{(u-m^2)} I_0(u) \labbel{eq:FI1}
\end{align}
where the second line is obtained fixing properly the boundary conditions.
It is very interesting to notice that all occurrences of integrals over 
elliptic integrals have cancelled out leaving space to a simple product 
of a logarithm and an elliptic integral.

\section{Derivation of all the relations at weight one } 
\labbel{sec:Eweight1all}\setcounter{equation}{0} 
\numberwithin{equation}{section} 
One might wonder whether the relation above is an accident or if, instead, 
such relations are more general.
It is not difficult to repeat the same exercise (i.e. deriving a second 
order differential equation, solving it, using integration by parts
and fixing the boundary conditions) for all the other weight-one possibilities.
Nevertheless, we find it more illuminating to follow a different (but of 
course equivalent) approach.\par
As we have seen, the operator $D(u,d/du)$ can be conveniently used to 
effectively reduce
the weight of the E-polylogarithms associated with the functions $I_0(u)$
and $J_0(u)$\footnote{We recall here that $I_0(u)$ and $J_0(u)$ solve a 
two by two system of differential equations together with $I_2(u)$ and $J_2(u)$.
The second order differential operator $D(u,d/du)$ can be used to lower the weight
of E-polylogarithms associated to $I_0(u)$ and $J_0(u)$, while
a different operator, say $D_2(u,d/du)$, should be used to lower the weight
of E-polylogarithms associated to $I_2(u)$ and $J_2(u)$.}. Following the example of generalized 
polylogarithms, we can therefore imagine to study the E-polylogarithms 
bottom-up, starting from weight one, and applying at each step the operator 
$D(u,d/du)$ to reduce the complexity to the previous weight, which can be 
considered as understood. \par

In order to see how this works, let us look again at the example above.
The function $\EI_0^{[1]}(0,u)$ is an E-polylogarithm of weight one.
From the discussion at the end of Section~\ref{sec:E1pol}, and 
in particular from Eq.(\ref{eq:GIn}), it is easy to see that similarly also 
the six functions 
\begin{align}
&f_1^{[1]}(u)=\ln{(u)} I_0(u)\,, \quad f_2^{[1]}(u)=\ln{(u-m^2)} I_0(u)\,, 
\quad  f_3^{[1]}(u)=\ln{(u-9m^2)} I_0(u)\,, \nonumber \\
&f_4^{[1]}(u)=\ln{(u)} J_0(u)\,, \quad f_5^{[1]}(u)=\ln{(u-m^2)} J_0(u)\,,
\quad  f_6^{[1]}(u)=\ln{(u-9m^2)} J_0(u)\, 
\,,
\end{align}
are E-polylogarithms of weight one.
It is then natural to consider the following linear combination
\be
A(u) = \EI_0^{[1]}(0,u) - \sum_{j=1}^6 c_j\; f_j^{[1]}(u) \labbel{eq:comb1}
\ee
where $c_j$ are constants. We can now apply the operator $D(u,d/du)$
on the function $A(u)$ and fix the coefficients $c_j$ such that 
\be
D\left( u, \frac{d}{du} \right) A(u) = 0\,.
\ee
By applying the operator $D(u,d/du)$ on each of the terms, we produce terms of 
weight zero, i.e. combinations of rational functions and $I_0(u)$, $J_0(u)$, 
$I_2(u)$, $J_2(u)$. 
By collecting for the independent terms and requiring the coefficients to be 
zero we find, as expected 
$$c_1=c_3=c_4=c_5=c_6=0 \quad \mbox{and} \quad c_2 = \frac{2}{3}\,.$$ 
This implies of course that 
\be \Du \left[ \EI_0^{[1]}(0,u) - \frac{2}{3} \ln(u-m^2)\ I_0(u) \right] 
    = 0 \ ,\labbel{eq:FE1} \ee 
and therefore, by Euler variation of the constants 
\be 
 \EI_0^{[1]}(0,u) - \frac{2}{3} \ln(u-m^2)\ I_0(u)  = \hat{c}_1\, I_0(u) 
 + \hat{c}_2\, J_0(u)\,, 
\ee 
where $\hat{c}_j$, $j=1,2$ are two numerical constants. 
By imposing the boundary conditions at $ u=9m^2 $ (according to 
Eq.(\ref{eq:J0at9}) $ J_0(u) $ has a logarithm singularity at that point, 
while all the other terms are regular, so that $ \hat{c}_2 = 0 $), 
we immediately find $\hat{c}_1 = \hat{c}_2 = 0\,,$ reproducing in this way 
the result in Eq.~\eqref{eq:FI1}.
\par      

In order to complete the exercise, we should remember that at order one we 
have two more functions to compute, namely $\EI_1^{[1]}(0,u)$ and 
$\EI_2^{[1]}(0,u)$. 
Clearly we see that, once $\EI_0^{[1]}(0,u)$ is known, then 
Eqs.(\ref{eq:udGl1u}, \ref{eq:Gl2}) allow us to compute 
$\EI_1^{[1]}(0,u)$ and $\EI_2^{[1]}(0,u)$.
In particular, $\EI_2^{[1]}(0,u)$ 
can be obtained from $\EI_0^{[1]}(0,u)$ 
by simple differentiation, while $\EI_1^{[1]}(0,u)$
fulfils a first order differential equation which can be solved by 
quadrature. Let us then proceed and compute them.
From Eq.~\eqref{eq:Gl2} we find immediately 
\begin{align} 
\EI_2^{[1]}(0,u) &= D_1\left( u, \frac{d}{du}\right) \EI_0^{[1]}(0,u) 
+ \left( -\frac{10}{3}m^2 \,u + m^4 + \frac{5}{9} u^2 \right) \EI_0^{[0]}(u) 
                                                             \nonumber \\ 
&+ \frac{2}{3}\left( 3m^2 +  u \right) \EI_1^{[0]}(u) - \EI_2^{[0]}(u) \,, 
\end{align} 
where the differential operator $D_1(u,d/du)$ is defined in 
Eq.~\eqref{eq:D1u}. Upon substituting Eq.~\eqref{eq:FI1} together with the 
weight zero results 
\be 
\EI_0^{[0]}(u) = I_0(u)\,, \quad \EI_1^{[0]}(u) = 0\,, 
                           \quad \EI_2^{[0]}(u) = I_2(u) 
\ee 
and working out the (straightforward) derivatives one finds easily 
\begin{align} 
\EI_2^{[1]}(0,u) = \frac{2}{3} \ln{(u-m^2)}  I_2(u)
  +   \frac{(u+3m^2)^2}{9} I_0(u) 
 - I_2(u)\,. \labbel{eq:Gl2res}
\end{align} 
\par 
Finally, let us consider $\EI_1^{[1]}(0,u)$. From Eq.(\ref{eq:udGl1u}), 
with $ b_i=4m^2 $, $ b_j=(W-m)^2 $ and $ u=W^2>9m^2$, we find 
\be \EI_1^{[1]}(0,u) = Il_1(4m^2,(W-m)^2,u) = 
    \int_{4m^2}^{(W-m)^2} \kern-15pt \frac{db}{\sqrt{R_4(u,b)}} 
        \left(b - \frac{u+3m^2}{3} \right) \ln{b}\ . 
\labbel{eq:Gl1u} \ee 
As 
\be \lim_{u\to 9m^2} \EI_1^{[1]}(0,u) = 0 \ , \labbel{eq:Gl1at9} \ee 
one has 
\be \EI_1^{[1]}(0,u) = \int_{9m^2}^u dv\ \frac{d}{dv}\EI_1^{[1]}(0,v) 
                                                \ ; \labbel{eq:Gl1dv} \ee 
from Eq.(\ref{eq:udGl1u}), recalling also Eq.(\ref{eq:valI1}), we have 
$$ \frac{d}{dv}\EI_1^{[1]}(0,v) = \biggl( \frac{2}{9} 
     - \frac{16m^2}{9(v-m^2)} \biggr) I_0(v) \ , $$ 
so that $ \EI_1^{[1]}(0,u) $ is given by the quadrature formula 
\be \EI_1^{[1]}(0,u) = \frac{2}{9}\, \int_{9m^2}^u dv\ \biggl( 1 
     - \frac{8\, m^2}{ v-m^2} \biggr) I_0(v) \ . \labbel{eq:Gl1dv1} \ee 
We are not able to simplify this expression further as we saw that the 
two integrals are linearly independent from each other, see 
Eq.~\eqref{eq:E1masters}.
We can nevertheless use Eq.s(\ref{eq:lnS}-\ref{eq:dlnU}), where $ S(u,b) $ 
and $ U(u,b) $ are defined, obtaining 
\be \EI_1^{[1]}(0,u) = \frac{2}{9} 
  \int_{4m^2}^{(W-m)^2} \kern-15pt \frac{db}{\sqrt{b(b-4m^2)}}  \ln{S(u,b)} 
+ \frac{4}{9} \int_{4m^2}^{(W-m)^2} \kern-15pt db 
  \left( \frac{1}{b} - \frac{1}{b-4m^2} \right) \ln{U(u,b)} \ . 
\labbel{eq:Gl1dv2} \ee 
We can now integrate by parts in $ b $ the last term of the above equation; 
by using the definition of $ \EI_1^{[1]}(0,u) $ Eq.(\ref{eq:Gl1u}) and the 
second of Eq.(\ref{eq:dlnU}) one finds the identity 
\be  \int_{4m^2}^{(W-m)^2} \kern-15pt \frac{db}{\rR} 
     \left(b - \frac{u+3m^2}{3} \right) 
     \left( \ln{b} + 2 \ln{(b-4m^2)} \right) = 
   \frac{2}{3} \int_{4m^2}^{(W-m)^2} \kern-15pt \frac{db}{\sqrt{b(b-4m^2)}} 
    \ln{S(u,b)} \ . \labbel{eq:idb} \ee 
\par 

\subsection{The relations at weight one}

Clearly, the procedure outlined above to compute $\EI_k^{[1]}(0,u)$, with 
$k=0,1,2$, can be easily repeated for all other weight-one functions 
$\EI_k^{[1]}(p_i,u)$, and for those involving the function $J_0(u)$, 
$\EJ_k^{[1]}(p_i,u)$.  
We proceed as follows
\begin{itemize}
\item[1-] First we use the second order differential operator $D(u,d/du)$ 
to determine relations between the functions $\EI_0^{[1]}(p_i,u)$, 
$\EJ_0^{[1]}(p_i,u)$ and the simpler products of logarithms with $I_0(u)$ 
and $J_0(u)$ functions, Eq.~\eqref{eq:comb1}. 
Surprisingly, at this order this allows us to rewrite all the functions of 
this form, where $p_i$ is on the the zeros in $ b $ of $R_4(u,b)$, as linear 
combinations of products of $I_0(u)$ or $J_0(u)$ and logarithms.

\item[2-] With this results at hand, we obtain the corresponding ones for the 
$\EI_2^{[1]}(p_i,u)$ and $\EJ_2^{[1]}(p_i,u)$ by differentiation.

\item[3-] Finally, we obtain an expression for the functions 
$\EI_1^{[1]}(p_i,u)$ and $\EJ_1^{[1]}(p_i,u)$ by integrating by quadrature 
their first order differential equation. 
\end{itemize}

We list here explicitly all the relations we find for the functions
$\EI_0^{[1]}(p_i,u)$ and $\EJ_0^{[1]}(p_i,u)$); for clarity 
we use the notation in terms of the $b$ integration. We find:

\be \int_0^{4m^2} \kern-10pt 
         \frac{db}{\sqrt{-R_4(u,b)}} \ \ln{b} 
       = \frac{2}{3}\ln(u-m^2)\ J_0(u) - \frac{4}{9}\pi I_0(u) \ ; 
\labbel{eq:FI1b} \ee 
\be \int_{4m^2}^{(W-m)^2} \kern-15pt 
         \frac{db}{\sqrt{R_4(u,b)}} \ \ln(b-4m^2) 
  = \left( \frac{1}{2}\ln(u-9m^2) + \frac{1}{6}\ln(u-m^2) \right) 
       I_0(u) - \frac{1}{2}\pi J_0(u) \ ; \labbel{eq:FI2a} \ee 
\be \int_0^{4m^2} \kern-10pt 
         \frac{db}{\sqrt{-R_4(u,b)}} \ \ln(b-4m^2) 
  = \left( \frac{1}{2}\ln(u-9m^2) + \frac{1}{6}\ln(u-m^2) \right) 
       J_0(u) - \frac{4}{9}\pi I_0(u) \ ; \labbel{eq:FI2b} \ee 
\begin{align} \int_{4m^2}^{(W-m)^2} \kern-15pt 
         \frac{db}{\sqrt{R_4(u,b)}} \ \ln((W-m)^2-b) &= 
  \left( \frac{1}{6}\ln(u-m^2) + \frac{1}{4}\ln{u} 
    + \frac{1}{2}\ln(W-m) + \frac{1}{2}\ln(W-3m) \right) I_0(u) \nonumber\\ 
       &- \frac{1}{2}\pi J_0(u) \ ; \labbel{eq:FI3a} \end{align} 
\begin{align} \int_0^{4m^2} \kern-10pt 
         \frac{db}{\sqrt{-R_4(u,b)}} \ \ln((W-m)^2-b) &= 
  \left( \frac{1}{6}\ln(u-m^2) + \frac{1}{4}\ln{u} 
    + \frac{1}{2}\ln(W-m) + \frac{1}{2}\ln(W-3m) \right) J_0(u) \nonumber\\ 
       &+ \frac{1}{18}\pi I_0(u) \ ; \labbel{eq:FI3b} \end{align} 
\begin{align} \int_{4m^2}^{(W-m)^2} \kern-15pt 
         \frac{db}{\sqrt{R_4(u,b)}} \ \ln((W+m)^2-b) &= 
  \biggl( \frac{1}{6}\ln(u-m^2) + \frac{1}{4}\ln{u} 
    + \frac{1}{2}\ln(W+m) + \frac{1}{2}\ln(W+3m) \biggr) I_0(u) \ ; 
                                  \nonumber\\  \labbel{eq:FI4a} \end{align} 
\begin{align} \int_0^{4m^2} \kern-10pt 
         \frac{db}{\sqrt{-R_4(u,b)}} \ \ln((W+m)^2-b) &= 
  \biggl( \frac{1}{6}\ln(u-m^2) + \frac{1}{4}\ln{u} 
     + \frac{1}{2}\ln(W+m) + \frac{1}{2}\ln(W+3m)  \biggr) J_0(u) 
     \nonumber\\ 
    &+ \frac{1}{18}\pi I_0(u) \ ; \labbel{eq:FI4b} \end{align}     
Note that if $u > 9m^2 $ all the appearing quantities are real, but the 
identities are of course valid in general if the proper analytic 
continuation is taken. For ease of typing, once more, we wrote 
$ \ln(b-4m^2), \ln(u-9m^2) $ {\it etc.} instead of 
$ \ln\left((b-4m^2)/m^2\right), \ln\left((u-9m^2)/m^2\right) $ .
We do not report here all the corresponding relations for the 
$\EI^{[1]}_2(p_i;u)$, $\EJ^{[1]}_2(p_i;u)$ and $\EI^{[1]}_1(p_i;u)$, 
$\EJ^{[1]}_1(p_i;u)$ for brevity, but it should be clear that they follow 
the same pattern as the ones for $\EI_2^{[1]}(0,u)$ and $\EI_1^{[1]}(0,u)$, 
derived respectively in Eqs.(\ref{eq:Gl2}, \ref{eq:Gl1dv1}). 
\newline 

Summarizing, the action of the differential operator $D(u,d/du)$ on the E-polylogarithms
of weight one associated to the functions $I_0(u)$ and $J_0(u)$
allows to reduce their weight and to determine algorithmically
surprising (and somewhat unexpected) relations
between E-polylogarithms and products of simple 
logarithms and the functions $I_0(u)$ and $J_0(u)$. 

\section{E-polylogarithms at weight two and beyond}
\labbel{sec:Eweight2}\setcounter{equation}{0} 
\numberwithin{equation}{section} 
The detailed study of the E-polylogarithms  at weight one revealed surprising identities
between the latter and products of complete elliptic integrals and
simple logarithms. We would like now to use similar methods to investigate
these functions at higher weights. We could of course repeat the same derivation
above, say, for the functions
$$\int_{b_i}^{b_j} \frac{db\, b^n}{ \sqrt{R_4(u,b)}} \, \ln^2{(b)}\,,$$
derive a second order differential equation satisfied by the latter, and solve it
by Eulers variation of constants. 

In order to have a better grasp of the general structure, nevertheless, it is useful
to study the more general class of functions defined by
\begin{equation}
I_\epsilon(n,u)  = \int_{4m^2}^{(W-m)^2} \; db \;
\frac{ b^n\; b^{\epsilon}  }
{\sqrt{R_4(u,b)}}\,.\labbel{eq:Gepsn}
\end{equation}

It is very easy to repeat the same procedure described above 
and show  that all these functions can be expressed in terms of
 three
independent master integrals, say
$$I_{\epsilon}(0,u) \,, 
\quad I_{\epsilon}(1,u) \,, 
\quad I_{\epsilon}(2,u)\,.$$
We can then perform the usual change of basis
\begin{align}
I_0(\epsilon,u) &= I_{\epsilon}(0,u) \nonumber \\
I_1(\epsilon,u) &= I_{\epsilon}(1,u) 
- \frac{(u + 3m^2)}{3}\,  I_{\epsilon}(0,u)\nonumber \\
I_2(\epsilon,u) &=I_{\epsilon}(2,u) 
- (u + 3m^2)\,  I_{\epsilon}(1,u)
- \frac{(u+3m^2)^2}{3} I_{\epsilon}(0,u)\,,
\end{align}
derive a system of differential equations satisfied by these functions, and turn
it into a second order differential equation for $I_0(\epsilon,u)$, together
with a first order differential equation for $I_1(\epsilon,u)$.
Note that in our notation we have
\begin{align}
I_j(\epsilon,u) = I_j(u) + \epsilon\;  \EI_j^{[1]}(0,u) +  \frac{1}{2}\, \epsilon^2 \; \EI_j^{[2]}(0,0,u)
+ \mathcal{O}(\epsilon^3) \quad \mbox{with} \quad j=0,1,2\,. \labbel{eq:exp}
\end{align}

The second order differential equation reads
\begin{align}
D\left( u, \frac{d}{du} \right) I_0(\epsilon,u)&= 
  \frac{2}{3}\, \epsilon \, \frac{(2u - 9m^2)}{u(u-m^2)(u-9m^2)} I_0(\epsilon,u)
  +   \frac{4}{3}\, \epsilon \;  \frac{1}{u-m^2} \,
\frac{d}{du} I_0(\epsilon,u)
\nonumber \\
&- \frac{1}{9} \, \epsilon^2\, \frac{(u-9m^2)^2}{u\,(u-m^2)^2} \, 
I_0(\epsilon,u)
- \frac{1}{3} \, \epsilon^2\;  \frac{1}{u\,(u-m^2)^2}\, I_1(\epsilon,u) 
\labbel{eq:IIordeps} \\
\end{align}
together with the equation for $I_1(\epsilon,u)$
\begin{align}
\frac{d}{du} I_1(\epsilon,u) &= \frac{2}{3} \, \epsilon\; 
\frac{ 1}{u-m^2} I_1(\epsilon,u) 
+ \frac{2}{9} \; \epsilon \; \frac{(u-9m^2)}{u-m^2} I_0(\epsilon; u) 
\labbel{eq:Sabryeps}
\end{align}
We see that there is a residual coupling (suppressed by two powers of 
$\epsilon$) 
between $I_0(\epsilon,u)$ and $I_1(\epsilon,u) $.
\par 
By expanding left- and right-hand-side of 
Eqs.(\ref{eq:IIordeps}, \ref{eq:Sabryeps})
and collecting for the terms proportional to $\epsilon^2$ 
we are left with the following equations

\begin{align}
D\left( u, \frac{d}{du} \right)  \EI_0^{[2]}(0,0,u)&= 
  \frac{4}{3}\,  \frac{(2u - 9m^2)}{u(u-m^2)(u-9m^2)}  \EI_0^{[1]}(0,u)
  +   \frac{8}{3}\, \frac{1}{u-m^2} \,
\frac{d}{du} \EI_0^{[1]}(0,u)
\nonumber \\
&- \frac{2}{9} \,  \frac{(u-9m^2)^2}{u\,(u-m^2)^2} \, 
 \EI_0^{[0]}(u)
- \frac{2}{3} \,  \frac{1}{u\,(u-m^2)^2}\,  \EI_1^{[0]}(u)
\end{align}
\begin{align}
\frac{d}{du} \EI_1^{[2]}(0,0,u) &= \frac{4}{3} \,  
\frac{ 1}{u-m^2} \EI_1^{[1]}(0,u) 
+ \frac{4}{9} \; \frac{(u-9m^2)}{u-m^2} \EI_0^{[1]}(0,u)\,,
\end{align}
while the results at previous orders read
\begin{align}
&\EI_0^{[0]}(u) = I_0(u)\,, \qquad \EI_0^{[1]}(0,u) = 
\frac{2}{3} \ln{(u-m^2)} I_0(u)\,, \nonumber \\
&\EI_1^{[0]}(u) = 0 \,, \quad 
\EI_1^{[1]}(0,u) = \frac{2}{9} \int_{9m^2}^u dv\,  \left[ 1 - \frac{8\, m^2}{v-m^2}  \right] I_0(v) \,.
\end{align}

Substituting all results explicitly and partial fractioning in $v$ we find
\begin{align}
\EI_0^{[2]}(0,0,u) &= c_1^{(2)} I_0(u) + c_2^{{2}} J_0(u) \nonumber \\
&+ \frac{16}{27\, \pi} \int^u dv \, 
F_{00}(u,v)\, \left( 4 + \frac{v}{m^2} + \frac{16 \,m^2}{v-m^2} \right) 
\ln{(v-m^2)}\,I_0(v)
\nonumber \\
&- \frac{16}{9\, \pi} \int^u dv \, F_{00}(u,v)\, \frac{ \ln{(v-m^2)} }{v-m^2} \,I_2(v)
\nonumber \\
&-\frac{16}{27\, \pi} \int^u dv \, 
F_{00}(u,v)\, \left( \frac{47}{4} - \frac{7}{4} \frac{v}{m^2}  
+ \frac{32\, m^2}{v-m^2} \right) \,I_0(v)\,, \labbel{eq:resulttwo}
\end{align}
and, since  $\lim_{u \to 9m^2} \EI_1^{[2]}(0,0,u) = 0$, 
\begin{align}
\EI_1^{[2]}(0,0,u) &= \frac{8}{27} \,  \int_{9m^2}^u  
\frac{ dv}{v-m^2} \int_{9m^2}^v dt\, \left[ 1 - \frac{8 \,m^2}{t-m^2}  \right] I_0(t)  \nonumber \\
&+ \frac{8}{27} \;\int_{9m^2}^u dv\, \left[ 1 - \frac{8\, m^2}{v-m^2}  \right]\ln{(v-m^2)} I_0(v)\,.
\labbel{eq:resultSabrytwo}
\end{align}

First of all, let us try to simplify Eq.~\eqref{eq:resultSabrytwo}.
Integrating by parts the first term in $dv$ we get at once
\begin{align}
\EI_1^{[2]}(0,0,u) &= \frac{8}{27} \,  
\ln{(u-m^2)} \int_{9m^2}^u dt\, \left[ 1 - \frac{8\, m^2}{t-m^2}  \right] I_0(t)  \nonumber \\
&- \frac{8}{27}\int_{9m^2}^u  \ln{(u-m^2)}  \left[ 1 - \frac{8\, m^2}{u-m^2}  \right] I_0(u) \nonumber \\
&+ \frac{8}{27} \;\int_{9m^2}^u dv\, \left[ 1 - \frac{8\, m^2}{v-m^2}  \right]\ln{(v-m^2)} I_0(v)\,\nonumber \\
&= \frac{8}{27}\, \ln{(u-m^2)} \int_{9m^2}^u dv\, \left[ 1 - \frac{8\, m^2}{v-m^2}  \right] I_0(v)\,,
\end{align}
where in the last line we renamed $t \to v$. Recalling the analytical 
result for $\EI_1^{[1]}(0,u)$ Eq.~\eqref{eq:Gl1dv1}, 
we see that 
we have
$$\EI_1^{[2]}(0,0,u) = \frac{4}{3} \ln{(u-m^2)} \, \EI_1^{[1]}(0,u)\,,$$
indeed, formally similar to the weight-one results for the
functions $\EI_0^{[1]}(0,u)$ and $\EI_2^{[1]}(0,u)$

Let us move now to Eq.~\eqref{eq:resulttwo} for $\EI_0^{[2]}(0,0,u)$.
At variance with order one, here we need to consider a more general class of
integrals
\begin{equation}\int^u\, dv 
\left\{ 1\;;\; v^n \;;\; \frac{1}{v^n} \;;\; \frac{1}{(v-m^2)^n} \,;\, \frac{1}{(v-9m^2)^n} \right\}
F_{0,0}(u,v)\,   \ln{(f(v))}  \, \Big\{ I_0(v) \, ; \, I_2(v) \Big\}\,, 
\end{equation}
where $f(v) = \left\{v, (v-m^2), (v-9 m^2) \right\}$.
Following the same logic as at weight one, we generate integration by parts identities
of the form
\begin{equation*}
\int^u\, dv \, \frac{d}{dv}\; \left(
 \left\{
1;\; v^n \;; \; \frac{1}{v^n} \;;\; \frac{1}{(v-m^2)^n} \;;\; \frac{1}{(v-9m^2)^n} \right\}
F_{0,0}(u,v)\,  \ln{ (f(v) )}  I_0(v) \right) = X_1(u)\,,
\end{equation*}
\begin{equation*}
\int^u\, dv \, \frac{d}{dv}\; \left(
 \left\{
1;\; v^n \;; \; \frac{1}{v^n} \;;\; \frac{1}{(v-m^2)^n} \;;\; \frac{1}{(v-9m^2)^n} \right\}
F_{0,0}(u,v)\,  \ln{(f(v))} I_2(v)  \right) = X_2(u)\,,
\end{equation*}
\begin{equation*}
\int^u\, dv \, \frac{d}{dv}\; \left(
 \left\{
1;\; v^n \;; \; \frac{1}{v^n} \;;\; \frac{1}{(v-m^2)^n} \;;\; \frac{1}{(v-9m^2)^n} \right\}
F_{0,2}(u,v)\,\ln{(f(v))} I_2(v)\right) = X_3(u)\,,
\end{equation*}
and solve the system of equations. Again we work with primitives, 
up to boundary terms, i.e. 
the functions $X_k(u)$ depend only on the variable $u$.
We find now that \textsl{for every choice} of logarithm, 
there are again 6 master integrals,
which we can choose once more as 
\begin{align}
\int^u\, dv \,   \left\{ 1\,; \,v\,; \,v^2\,;\, \frac{1}{v}\, ;\, \frac{1}{v-m^2} \,;\, \frac{1}{v-9 m^2}  \right\} \,
F_{0,0}(u,v)\, \ln{(f(v))} \, I_0(v)\,
\end{align}
for $f(v) = \left\{v, (v-m^2), (v-9 m^2) \right\}$\,.
More explicitly, once again we find that one of the integrals in 
Eq.~\eqref{eq:resulttwo} 
can be expressed as linear combination of the others as follows
\begin{align}
 \int^u dv \, F_{00}(u,v)\, \frac{ \ln{(v-m^2)} }{v-m^2} \,I_2(v) &=
 \frac{1}{3}   \int^u dv \,   \left( 4 m^2 + v + \frac{16 \, m^4}{v-m^2} \right)  \ln{(v-m^2)}\, F_{0,0}(u,v)\,I_0(v)\nonumber \\
 &- \frac{1}{3}  \int^u dv \,  \left( 8 m^2 - v + \frac{8\, m^4}{v-m^2} \right) \, 
 F_{0,0}(u,v)\,I_0(v) \nonumber 
\\ &- \frac{\pi}{2} I_0(u)\,   \int^u dv \, \frac{\ln{(v-m^2)}}{v-m^2}\,.
\end{align}
Using this identity in Eq.~\eqref{eq:resulttwo} we see that 
the highest weight do cancel, similarly to the previous order, and
we are left with
\begin{align}
\EI_0^{[2]}(0,0,u) &= \int_{4m^2}^{(W-m)^2} \frac{db}{\sqrt{R_4(u,b)}} \ln^2{b}  
= c_1^{(2)} I_0(u) + c_2^{(2)} J_0(u) \nonumber \\
&+ \frac{4}{9} I_0(u) \ln^2{(u-m^2)}
- \frac{4}{9\, \pi} \int_{9m^2}^u dv \, F_{0,0}(u,v)\, 
\left( 5m^2 -  v + \frac{32\, m^4}{v-m^2} \right) \,I_0(v)
\nonumber \\
&= \frac{4}{9} I_0(u) \ln^2{(u-m^2)}
- \frac{4}{9\, \pi} \int_{9m^2}^u dv \, F_{0,0}(u,v)\, 
\left( 5m^2 -  v + \frac{32\, m^4}{v-m^2} \right) \,I_0(v)\,,
\labbel{eq:resulttwo2}
\end{align}
where in the last line we fixed the boundary conditions 
finding $c_1^{(2)} = c_2^{(2)} = 0$\,.

The result in Eq.~\eqref{eq:resulttwo2} shows interesting features.
Indeed, differently from the weight-one case, not all integrals over 
the functions $F_{0,0}(u,v)$ have disappeared. Nevertheless,
 we see that the piece of highest transcendental weight, i.e. the one involving integrals over
 $F_{0,0}(u,v)$ and logarithms in this case, can indeed be eliminated in favour of
a simpler term which contains a logarithm squared multiplied by $I_0(u)$.
The remaining integrals are simpler, as they do not contain any logarithms.
\subsection{Relations for E-polylogarithms at weight two}
Having discussed explicitly the case with a $\ln^2{b}$, we can now 
in principle study all other weight-two E-polylogarithms, 
including possibly those containing di-logarithms $\Li_2(f(b))$ with 
branches corresponding to the roots of the polynomial $R_4(u,b)$.
We can do this similarly to weight one, namely writing a general Ansatz
and using the second order differential operator $D(u, d/du)$ to fix
the coefficients.

As exemplification, let us consider the following weight two E-polylogarithms
\begin{align}
E_0(u) &=\int_{4 m^2}^{(W-m)^2} \frac{db}{\rR} \,  \ln^2{ b } \nonumber \\
E_1(u) &= \int_{4 m^2}^{(W-m)^2} \frac{db}{\sqrt{R_4(u,b)}} \,  \ln^2{ (b-4m^2) } \nonumber \\
E_2(u) &= \int_{4 m^2}^{(W-m)^2} \frac{db}{\sqrt{R_4(u,b)}} \,  
\left[ \ln{ (b-(W-m)^2) } + \ln{ (b-(W+m)^2) } \right]^2 \nonumber \\ 
E_3(u) &= \int_{4 m^2}^{(W-m)^2} \frac{db}{\sqrt{R_4(u,b)}} \, \ln{b} \ln{ (b-4m^2) } \nonumber \\
E_4(u) &= \int_{4 m^2}^{(W-m)^2} \frac{db}{\sqrt{R_4(u,b)}} \, \ln{b}\left[ \ln{ (b-(W-m)^2) } + \ln{ (b-(W+m)^2) } \right]
\nonumber \\
E_5(u) &= \int_{4 m^2}^{(W-m)^2} \frac{db}{\sqrt{R_4(u,b)}} \, \ln{(b-4m^2)}\left[ \ln{ (b-(W-m)^2) } + \ln{ (b-(W+m)^2) } \right]\,. \labbel{eq:Eweight2}
\end{align} 
All these functions can be rewritten in the notation $\EI_0^{[2]}(p_i,p_j;u)$, up to analytic continuation.
This is achieved by simply rewriting the (products) of logarithms as standard multiple-polylogarithms,
for example
\be
\ln{(b-4m^2)} = G(4m^2, b) + \ln{4m^2} \pm i \,\pi\,,
\ee
depending on the imaginary part given to $b$.
We use here a standard representation in terms of logarithms to keep the formulas
as clear as possible.

In order to build an Ansatz that is large enough to match all these functions, we should
consider all functions that behave as weight two or one E-polylogarithms under the 
action of the operator $D(u,d/du)$. 
First of all, we include the simplest E-polylogarithms, obtained by multiplying $I_0(u)$
or $J_0(u)$ by standard multiple polylogarithms.
\begin{align}
A(u) &= \left[ a_0 \ln{u} + a_1 \, \ln{(u-m^2)} + a_2 \, \ln{(u-9m^2)} \right] I_0(u) \nonumber \\
&+ \left[a_3 \ln^2{u} + a_4 \, \ln^2{(u-m^2)} + a_5 \, \ln^2{(u-9m^2)} \right] I_0(u) \nonumber \\
&+ \left[ a_6 \ln{u} \ln{(u-m^2)} + a_7 \, \ln{u} \ln{(u-9m^2)} + a_8 \, \ln{(u-m^2)} \ln{(u-9m^2)} \right] I_0(u)
\nonumber \\
&+\left[ b_0 \ln{u} + b_1 \, \ln{(u-m^2)} + b_2 \, \ln{(u-9m^2)} \right] J_0(u) \nonumber \\
&+ \left[b_3 \ln^2{u} + b_4 \, \ln^2{(u-m^2)} + b_5 \, \ln^2{(u-9m^2)} \right] J_0(u) \nonumber \\
&+ \left[ b_6 \ln{u} \ln{(u-m^2)} + b_7 \, \ln{u} \ln{(u-9m^2)} + b_8 \, \ln{(u-m^2)} \ln{(u-9m^2)} \right] J_0(u)\,,
\labbel{eq:Ansatz1}
\end{align}
where the $a_j$ and $b_j$ are numerical coefficients.
Note that here, for simplicity, we did not include  di-logarithms, which in a more general
case should also be included.  Simple (products of) logarithms seem to be enough
as long as we limit ourselves to (products of) logarithms in the functions~\eqref{eq:Eweight2}.
We have verified explicitly that allowing for the presence of a  di-logarithm 
under the integration sign, requires also to enlarge the Ansatz Eq.~\eqref{eq:Ansatz1}
allowing for di-logarithms as well. We do not report these results
for brevity.

The Ansatz Eq.~\eqref{eq:Ansatz1} is not complete, as we can see from 
the explicit result in Eq.~\eqref{eq:resulttwo2}.
From the discussion in Section~\ref{sec:FF}, it is clear that, in general,
we must include in the Ansatz 6 more functions, i.e. the master integrals
in Eq.~\eqref{eq:Mastersw1}. We write therefore 

\begin{align}
A_{tot}(u) &= A(u) + c_1 \int^u dv F_{0,0}(u,v) I_0(v) 
 + c_2 \int^u \frac{dv}{v} F_{0,0}(u,v) I_0(v) \nonumber \\
& + c_3 \int^u \frac{dv}{v-m^2} F_{0,0}(u,v) I_0(v) 
 + c_4 \int^u \frac{dv}{v-9m^2} F_{0,0}(u,v) I_0(v)  \nonumber \\
& +  c_5 \int^u dv\, v F_{0,0}(u,v) I_0(v) 
 + c_6 \int^u  dv\, v^2 F_{0,0}(u,v) I_0(v)\,.
\labbel{eq:Ansatztot}
\end{align}

We act with $D(u,d/du)$ on the combination $(E_i(u) - A_{tot}(u))$ for every $i=1,...,5$,
use the results at weight one and
collect for the independent structures. 
Imposing $$\Du \left[ E_i(u) - A_{tot}(u) \right] = 0\,,$$
we obtain a linear system for the coefficients
of the Ansatz which we can solve straightforwardly. 
This fixes the result uniquely up to boundary terms. We find 

\begin{align}
\overline{E}_1(u) &= \frac{1}{36}\, \left[ \ln{(u-m^2)} + 3 \ln{(u-9m^2)} \right]^2 I_0(u)
- \frac{\pi}{6} \left[ \ln{(u-m^2)} + 3 \ln{(u-9m^2)} \right] \, J_0(u) \nonumber \\
&- \frac{4}{9\, \pi} \int^u dv F_{0,0}(u,v)\, \left[ 3 m^2 -  \frac{6\,m^4}{v-m^2} + \frac{54\, m^4}{v-9m^2} + v\right] I_0(v) \nonumber \\ & \nonumber \\
\overline{E}_2(u) &= \frac{1}{36}\, \left[ 3 \ln{u} + 5 \ln{(u-m^2)} + 3 \ln{(u-9m^2)} \right]^2 I_0(u)\nonumber \\
&- \frac{\pi}{6} \left[ 3 \ln{u} + 5 \ln{(u-m^2)} + 3 \ln{(u-9m^2)} \right] \, J_0(u) \nonumber \\
&- \frac{1}{9\, \pi} \int^u dv F_{0,0}(u,v)\, \left[ 2 m^2 + \frac{27\,m^4}{v} 
- \frac{88\,m^4}{v-m^2} + \frac{216\, m^4}{v-9m^2} + 3\, v\right] I_0(v) \nonumber \\ & \nonumber \\
\overline{E}_3(u) &= \frac{1}{9}\, \ln{(u-m^2)}\,  \left[  \ln{(u-m^2)} + 3 \ln{(u-9m^2)} \right] I_0(u)
- \frac{\pi}{3} \,  \ln{(u-m^2)} \, J_0(u)  \nonumber \\ \displaybreak
&+ \frac{4}{9\, \pi} \int^u dv F_{0,0}(u,v)\, \left[ m^2 -  \frac{8\,m^4}{v-m^2} -  v\right] I_0(v) \nonumber \\ & \nonumber \\
\overline{E}_4(u) &=  \frac{1}{9}\, \ln{(u-m^2)}\,  \left[  3\ln{u} + 5\ln{(u-m^2)} + 3 \ln{(u-9m^2)} \right] I_0(u)
- \frac{\pi}{3} \,  \ln{(u-m^2)} \, J_0(u) \nonumber \\
&+ \frac{2}{9\, \pi} \int^u dv F_{0,0}(u,v)\, \left[ 3 m^2 + \frac{48\,m^4}{v-m^2} + v\right] I_0(v) \nonumber \\ & \nonumber \\
\overline{E}_5(u) &=  \frac{1}{36}\, \left[ \ln{(u-m^2)} + 3 \ln{(u-9m^2)}\, \right]  \left[  3\ln{u} + 5\ln{(u-m^2)} + 3 \ln{(u-9m^2)} \right] I_0(u) \nonumber \\
&- \frac{\pi}{4} \,  \left[ \ln{u} + 2\, \ln{(u-m^2)} + 2\,\ln{(u-9m^2)}   \right] \, J_0(u) \nonumber \\
&- \frac{2}{9\, \pi} \int^u dv F_{0,0}(u,v)\, \left[ 13 m^2 + \frac{4\,m^4}{v-m^2} 
+ \frac{108\,m^4}{v-9m^2} - v\right] I_0(v)\,, \labbel{eq:Funnyweight2}
\end{align}
with 
$$E_k(u) = \overline{E}_k(u) + c_{1}^{(k)}\, I_0(u) + c_{2}^{(k)}\, J_0(u)\,,
 \qquad \mbox{for}\quad k=1,...,5\,.$$

\section{The imaginary part of the two-loop massive sunrise} 
\labbel{sec:sun}\setcounter{equation}{0} 
\numberwithin{equation}{section} 
As we announced in the introduction, a subset of the class of functions
analyzed here is of direct physical interest for the computation of the 
(imaginary part of the) two-loop
massive sunrise graph.
In fact, up to an irrelevant multiplicative phase, in $d = (2-2 \epsilon)$ 
dimensions it is well known that~\cite{Remiddi:2016gno}
\be \Ima S_{\epsilon}(u) = \Ima \,\Biggl[\  \sunrise{p} \ \ \Biggr] = 
   \int_{4m^2}^{(W-m)^2} \frac{db}{\rR} 
          \left( \frac{R_4(u,b)}{ub} \right)^{-\epsilon} 
\labbel{eq:ImSuneps} \ee 
with $p^2 = u$ and $W=\sqrt{u}$. Clearly, upon expanding in $\epsilon$ up 
to $\epsilon^2$ we find 
\begin{align}
u^{-\epsilon} \, \Ima S_{\epsilon}(u) = I_0(u) 
&+ \epsilon \int_{4m^2}^{(W-m)^2} db 
\frac{ \left[ \ln{(b-4m^2)} + \ln{((W-m)^2 - b)} + \ln{((W+m)^2 - b)}\right]}
{\sqrt{[R_4(u,b)]}} \nonumber \\
&+ \frac{1}{2}\,\epsilon^2 \, \left[ E_0(u) + E_2(u) + 2 E_5(u)\right] 
+ \mathcal{O}(\epsilon^3)\,,
\end{align}
where the functions $ E_i(u) $ are defined in~\eqref{eq:Eweight2}.
Using the results obtained at weight one Eqs.(\ref{eq:FI1}, \ref{eq:FI2a}, 
\ref{eq:FI3a}, \ref{eq:FI4a}) and at weight 2 Eqs.~\eqref{eq:Funnyweight2}, 
we find finally (again up to a boundary condition)
\begin{align}
\Ima S_{\epsilon}(u) &= I_0(u) 
+ \epsilon \left[ \pi J_0(u) - \frac{1}{2} \, l(u)\, I_0(u) \right] 
+ \epsilon^2 \, \left[ \frac{1}{8} l^2(u) I_0(u) - \frac{\pi}{2} \, l(u) \,J_0(u) \right. \nonumber \\
& \left. - \frac{1}{\pi}\, \int^u dv F_{0,0}(u,v)\, \left( \frac{11\, m^2 }{3} + \frac{3\, m^4 }{2\, v} 
- \frac{16\,m^4}{3(v-m^2)} 
+ \frac{48\,m^4}{v-9m^2} + \frac{v}{6} \right) I_0(v)
\right] + \mathcal{O}(\epsilon^3)\,,
\end{align}
where we introduced the shorthand notation
$$l(u) = 2 \ln{(u-m^2)} + 2 \ln{(u-9m^2)} - \ln{u} \,.$$

\section{Conclusions}
\labbel{sec:conc}
In this paper we studied a class of functions, dubbed E-polylogarithms, 
which constitutes a
natural elliptic generalization of multiple polylogarithms.
A subset of the functions analyzed here is relevant for the calculation
of the imaginary part of the two-loop massive sunrise graph.
\par
While standard polylogarithms fulfil simple first order differential equations
with rational coefficients, we showed that E-polylogarithms 
fulfil a system of three by three linear first order differential equations, 
which can be decoupled in a two by two coupled system, 
plus a decoupled first order 
differential equation. These equations can be solved by Euler's variation
of constants, providing a representation of these functions as iterated integrals
over rational factors and products of complete elliptic integrals.
\par
This allows to tentatively associate to the E-polylogarithms a weight, dubbed E-weight, 
which turns out to be naturally lowered by the action of the corresponding 
(matricial or higher order) differential operator. In this way we could study properties and
relations among E-polylogarithms bottom-up in their E-weight 
and show, in particular, that all E-weight one E-polylogarithms 
can be rewritten as products of standard
polylogarithms and complete elliptic integrals.
Starting at E-weight equal to two, this is not true anymore and E-polylogarithms
introduce genuine new structures. Nevertheless, also at E-weight two,
we found interesting relations for the highest transcendental piece of 
the E-polylogarithms in terms of products of weight-two standard
polylogarithms and complete elliptic integrals. 
Finally, we used these results to provide a compact representation
for the order $\epsilon^2$ of the imaginary part of the 
two-loop massive sunrise graph.
\par
While our study is not definitive, it might open interesting possibilities
for the systematic study and simplification of functions appearing in the
calculation of multiloop Feynman graphs with many scales and/or
massive propagators. 
Indeed, the analytic calculation of Feynman integrals
which fulfil higher order differential equations still remains largely
out of reach; a first obstruction was given by the absence of a systematic understanding
of the solution of their corresponding higher-order homogeneous equations.
Quite recently it was shown that the study of the maximal cut of Feynman
integrals provides an efficient tool to determine the missing homogeneous
solutions~\cite{Primo:2016ebd,Frellesvig:2017aai,Bosma:2017ens,Primo:2017ipr,Harley:2017qut}
and this obstruction was partially lifted.\footnote{For interesting developments on the
relation between unitarity cuts and the analytic properties of Feynman
integrals see for 
example~\cite{CaronHuot:2012ab,Huang:2013kh,Sogaard:2014ila,Hauenstein:2014mda,
Abreu:2017ptx,Abreu:2017enx,Abreu:2017mtm}.}
Thanks to these developments, in fact, we are now in the position to systematically write
integral representations for the solutions of complicated Feynman integrals;
the crucial problem remains therefore that of studying the properties of
these functions and of the relations among them, which is one of the most 
important aspect of an analytic calculation.
\par
The methods described in this paper are, at least in principle, 
not limited to elliptic generalizations 
of multiple polylogarithms and can instead be equally well applied to the
study of functions which fulfil even higher order
differential equations. We hope therefore that they can be of
some use for a systematic
analysis of the properties of Feynman integrals beyond multiple
polylogarithms.

\section*{Acknowledgements} 
We are grateful to J. Vermaseren for his assistance in the use of the 
algebraic program FORM~\cite{Vermaseren:2000nd} which was intensively used 
in all the steps of this work. 
We acknowledge very interesting discussions with Claude Duhr on the properties of 
elliptic generalizations of multiple polylogarithms.

\noindent
The authors thank the Munich Institute for Astro- and Particle Physics (MIAPP) of the DFG cluster of excellence ``Origin and Structure of the Universe" for support, where this
work was completed.

\appendix 

\section{The analytical calculation of four master integrals} 
\labbel{App:Masters}
Concerning the analytic expressions of the {\it master integrals} 
given in~\eqref{eq:E1masters} 
an explicit calculation, obtained by using the integral representation 
Eq.(\ref{eq:I0u}) and by exchanging the order of integration, gives 
\begin{align} \int_{9m^2}^u dv\ I_0(v) &= \int_{9m^2}^u dv 
   \int_{4m^2}^{(\sqrt{v}-m)^2} \kern-15pt \frac{db}{\sqrt{R_4(v,b)}}  
                                                                  \nonumber\\ 
   &= \int_{4m^2}^{(W-m)^2} \kern-15pt \frac{db}{\sqrt{b(b-4m^2)}} 
   \int_{(\rb+m)^2}^u \frac{dv}{\sqrt{R_2(v,b,m^2)}} \nonumber\\ 
 &= \int_{4m^2}^{(W-m)^2} \kern-15pt \frac{db}{\sqrt{b(b-4m^2)}} 
   \ln{S(u,b)} \ , \labbel{eq:lnS} \end{align} 
where 
\begin{align} 
   R_4(v,b) &= b(b-4m^2) R_2(v,b,m^2) \ , \nonumber\\  
   R_2(v,b,m^2) &= (v-(\rb-m)^2)(v-(\rb+m)^2) \ , \labbel{eq:R2vb} 
\end{align} 
and 
\be {S(u,b)} = \frac{ \sqrt{u-(\rb-m)^2} + \sqrt{u-(\rb+m)^2} } 
                    { \sqrt{u-(\rb-m)^2} - \sqrt{u-(\rb+m)^2} } 
\ , \labbel{eq:Sub} \ee 
so that the derivatives of $  \ln{S(u,b)} $ are 
\begin{align} 
  \frac{d}{du} \ln{S(u,b)} &= \frac{1}{\sqrt{R_2(u,b,m^2)}} 
                                           \ , \nonumber\\ 
  \frac{d}{db} \ln{S(u,b)} &= -\ \frac{u+b-m^2}{2b\sqrt{R_2(u,b,m^2)}} 
                                           \ . \labbel{eq:dlnS} 
\end{align} 
One finds, similarly 
\be \int_{9m^2}^u dv\ \frac{1}{v}\ I_0(v) = 
    \int_{4m^2}^{(W-m)^2} \kern-15pt \frac{db}{\sqrt{b(b-4m^2)}} 
    \frac{1}{b-m^2} \ln{T(u,b)} \ , \labbel{eq:lnT} \ee 
where 
\begin{align} 
    T(u,b) &= \frac{ (\rb+m)\sqrt{u-(\rb-m)^2} + (\rb-m)\sqrt{u-(\rb+m)^2} } 
                   { (\rb+m)\sqrt{u-(\rb-m)^2} - (\rb-m)\sqrt{u-(\rb+m)^2} } 
              \ , \nonumber\\ 
   \frac{d}{du} \ln{T(u,b)} &= \frac{b-m^2}{u\sqrt{R_2(u,b,m^2)}} 
              \ , \nonumber\\ 
   \frac{d}{db} \ln{T(u,b)} &= \frac{u-3b-m^2}{2b\sqrt{R_2(u,b,m^2)}} \ , 
\labbel{eq:dlnT} \end{align} 
and 
\be 
   \int_{9m^2}^u \frac{dv}{v-m^2}\ I_0(v) = 
   \int_{4m^2}^{(W-m)^2} \kern-15pt \frac{db}{b(b-4m^2)} \ln{U(u,b)} \ , 
\labbel{eq:lnU} \ee 
with 
\begin{align} 
 U(u,b) &= \frac{   \sqrt{\rb+2m}\sqrt{u-(\rb-m)^2} 
                  + \sqrt{\rb-2m}\sqrt{u-(\rb+m)^2} } 
                {   \sqrt{\rb+2m}\sqrt{u-(\rb-m)^2} 
                  - \sqrt{\rb-2m}\sqrt{u-(\rb+m)^2} } \ , \nonumber\\ 
 U^2(u,b) &= \frac{ (u-b+3m^2)b + \rR } 
                  { (u-b+3m^2)b - \rR } \ , \nonumber\\ 
  \frac{d}{du} \ln{U(u,b)} &= \frac{b(b-4m^2)}{(u-m^2) \rR } 
                                                      \ , \nonumber\\ 
  \frac{d}{db} \ln{U(u,b)} &= \frac{u-3b+3m^2}{2 \rR } \ . 
\labbel{eq:dlnU} \end{align} 
For the master integral with the factor $ 1/(v-9m^2) $ it is convenient 
to integrate in $ v $ in the region $ u_0 < v < u $, with $ u_0 > 9m^2 $, 
as the limit $ u_0 \to 9m^2 $ is logarithmically divergent. 
One finds  
\begin{align} 
   \int^u \frac{dv}{v-9m^2}\ I_0(v) 
   &= \int_{4m^2}^{(\ruz-m)^2} 
    \frac{db}{\sqrt{b(b-16m^2)}}\ \frac{1}{b-4m^2} 
    \ \ln\frac{V(u,b)}{V(u_0,b)} \nonumber\\ 
   &+ \int_{(\ruz-m)^2}^{(\ru-m)^2} 
    \frac{db}{\sqrt{b(b-16m^2)}}\ \frac{1}{b-4m^2}\ \ln{V(u,b)} \ , 
\labbel{eq:lnV} \end{align} 
with 
\begin{align} 
  V(u,b) &= \frac{   \sqrt{(\rb+4m)(\rb-2m)} \sqrt{u-(\rb-m)^2} 
                   + \sqrt{(\rb-4m)(\rb+2m)} \sqrt{u-(\rb+m)^2} } 
                 {   \sqrt{(\rb+4m)(\rb-2m)} \sqrt{u-(\rb-m)^2} 
                   - \sqrt{(\rb-4m)(\rb+2m)} \sqrt{u-(\rb+m)^2} } 
          \ , \nonumber\\ &\nonumber \\
  \frac{d}{du} \ln{V(u,b)} &= \frac{ \sqrt{(b-4m^2)(b-16m^2)} } 
         { (v-9m^2)\sqrt{R_2(u,b,m^2)}} \ , \nonumber\\ 
  \frac{d}{db} \ln{V(u,b)} &= 
  \frac{-3 b^2+m^2 (11 b+8 u)+b u-8 m^4}{2 b \sqrt{(b-4m^2)(b-16m^2)}
   \sqrt{R_2(u,b,m^2)}}
\ . \labbel{eq:dlnV} 
\end{align} 
(Note that the integrand in the {\it r.h.s.} of (\ref{eq:lnV}) is real, 
even if some square roots are imaginary when $ \rb < 4m $). 
\par 

\section{Another integral representation for $ I_0(u) $} 
\labbel{sec:fext} \setcounter{equation}{0} 
\numberwithin{equation}{section} 
As an extension of the procedure outlined in Section~\ref{sec:Thebeg}, 
we will derive a second order equation for the integral 
\be I((W+m)^2,\infty,u) = \int_{(W+m)^2}^{\infty}  
                       \frac{db}{\sqrt{R_4(u,b)}} \ , \labbel{eq:I+oo} \ee 
with $u $ in the range  $ 9m^2 < u < \infty $ for definiteness. The 
integral is convergent, but we cannot follow exactly all the steps of 
Section~\ref{sec:Thebeg}, as the direct use of Eq.(\ref{eq:I(n,u)}), for 
instance, would involve meaningless (non-convergent) integrals like 
$$  \int_{(W+m)^2}^{\infty} 
                       \frac{db}{\sqrt{R_4(u,b)}} b^n \ , $$ 
with integer positive $ n $. \par 
In order to follow as much as possible the procedure leading to 
Eq.(\ref{eq:Eq2}), we introduce instead the quantities 
\be I((W+m)^2,B,n,u) = \int_{(W+m)^2}^B 
                    \frac{db}{\sqrt{R_4(u,b)}} b^n\ , \labbel{eq:IbnB} \ee 
where $ B $ is a parameter satisfying the condition $ B \gg u $, and 
correspondingly modify the (homogeneous) identities Eq.(\ref{eq:dbbd}) into 
the (inhomogeneous) relations 
\be \int_{(W+m)^2}^B db\ \frac{d}{db}\left(\sqrt{R_4(u,b)}\ b^n \right) 
            =  \sqrt{R_4(u,B)}\ B^n \ .    \labbel{eq:dbbdB} \ee 
From it, one obtains an equation whose {\it l.h.s.} is identical to the 
{\it l.h.s.} of Eq.(\ref{eq:idbn}), homogeneous in the 
quantities $ I((W+m)^2,B,n,u) $, while the {\it r.h.s.}, which is not zero, 
can be considered as an inhomogeneous term, depending on $ u $ and the 
parameter $ B $ only, but not on the quantities $ I((W+m)^2,B,n,u) $. 
From this point on we can follow closely the derivation 
of Section~\ref{sec:Thebeg}, introducing as there all the auxiliary quantities 
related to the original $ I((W+m)^2,B,n,u) $ and obtaining at each step 
relations which have the same homogeneous part and, in addition, a few non 
vanishing inhomogeneous terms depending on $ B $. \par 
As a result, instead of Eq.(\ref{eq:Eq2}) we get 
$$ \Du \ I_0((W+m)^2,B,u) = - \frac{1}{uB^2}( 1 + \cdots ) \ , $$ 
where the dots stand for terms of higher order in $ 1/B $. In the 
$ B \to \infty $ limit, the equation becomes 
\be \Du \ I((W+m)^2,\infty,u) = 0 \ , \labbel{eq:Eq2oo} \ee 
identical to Eq.(\ref{eq:Eq2}), showing that the function defined by 
Eq.(\ref{eq:I+oo}) is also a solution of Eq.(\ref{eq:DuI}). \par 
We have already observed that all the solutions of Eq.(\ref{eq:DuI}) are 
linear combinations of $ I_0(u) $, Eq.(\ref{eq:I0u}), 
and $ J_0(u) $, Eq.(\ref{eq:J0u}); as an elementary calculation gives 
$$ \lim_{u\to 9m^2}I((W+m)^2,\infty,u) = \frac{\sqrt{3}}{36m^2}\pi 
                                           \ , $$ 
for comparison with Eq.s(\ref{eq:I0at9}),(\ref{eq:J0at9}) one has 
\be I((W+m)^2,\infty,u) = \frac{1}{3}\ I_0(u) \ . \labbel{eq:Ioo=I0} \ee 
\par 
In the same way, one finds that also the function 
\be I(-\infty,0,u) = \int_{-\infty}^{0} 
                       \frac{db}{\sqrt{R_4(u,b)}} \ , \labbel{eq:I-oo} \ee
with $u $ in the range  $ 9m^2 < u < \infty $, is another solution of 
Eq.(\ref{eq:Eq2}). From its value at $ u=9m^2 $ one finds 
\be I(-\infty,0,u) = \frac{2}{3}\ I_0(u) \ .       \labbel{eq:I-oo=I0} \ee 
Without entering into further details, let us just observe that by contour 
integration arguments in the complex $ b $ plane one can obtain 
the relation 
\be I(-\infty,0,u) + I((W+m)^2,\infty,u) = I_0(u) \ , \labbel{eq:ooI0oo} 
\ee 
which involves the sum of Eq.s(\ref{eq:Ioo=I0}) and (\ref{eq:I-oo=I0}), 
but not the two quantities separately. 

\section{The relation between $I(1,u)$ and $I(0,u)$} 
\labbel{sec:AppX} \setcounter{equation}{0} 
\numberwithin{equation}{section} 
We comment here briefly Eq.(\ref{eq:udI1u}), whose content is 
\be \frac{d}{du}\int_{b_i}^{b_j}\frac{db}{\sqrt{R_4(u,b)}}
         \left(b - \frac{u+3m^2}{3} \right) = 0 \ . \nonumber     \ee 
To our knowledge, that result was found in 1962 by A.Sabry~\cite{Sabry}, 
albeit in a somewhat different notation, see Eq.(88) of~\cite{Sabry}, 
for the particular case $ b_i = 4m^2, b_j = (W-m)^2 $, and used 
to derive Eq.(85) of that paper, which in our notation reads 
\be I(4m^2,(W-m)^2,1,u) - \frac{u+3m^2}{3}I(4m^2,(W-m)^2,0,u) = 0 . 
\labbel{eq:Sabry} \ee 
The result was independently reobtained in~\cite{Laporta:2004rb}, 
see the derivation of Eq.(7.7) there (and later repeated in Eq.s(A.8,9,10) 
of~\cite{Remiddi:2016gno}) by using the relation 
\be \frac{d}{du}\ln\frac{(u-b+3m^2)b+\rR}{(u-b+3m^2)b-\rR} 
   = \frac{u-3b+3m^2}{\rR} \ , \labbel{eq:SabryLap} \ee 
and writing 
\be \int_{b_i}^{b_j} \kern-5pt 
   db\ \frac{3b-u-3m^2}{\sqrt{R_4(u,b)}} = 
   \int_{b_i}^{b_j} db \frac{d}{db} \left( 
  \ln{\frac{b(u-b+3m^2)+\sqrt{R_4(u,b)}}{b(u-b+3m^2)-\sqrt{R_4(u,b)}}} 
  \right) \ . \labbel{eq:C.3} \ee 
If $ b_i = 4m^2, b_j = (W-m)^2 $ the logarithm vanishes at the end points 
of the integration and Eq.\eqref{eq:Sabry} is recovered. \par 
Eq.\eqref{eq:SabryLap}, fully equivalent to Eq.\eqref{eq:dlnU} of the 
present paper, was already given in~\cite{Barbieri:1974nc}, 
just after Eq.(5.8) there (but unfortunately with typing errors!).  
\par 
As explained in~\cite{Remiddi:2016gno}, if in~\eqref{eq:C.3} the end points 
of the integration are taken to be a different pair of roots of the polynomial 
$ R_4(u,b) $, one can have a non vanishing result; indeed, for $b_1=0$ 
and $b_2=4m^2$ one finds 
\be \int_0^{4m^2} \frac{db}{\sqrt{-R_4(u,b)}}\left(b-\frac{u+3m^2}{3} 
                 \right) = - \frac{1}{3}\pi \ , \labbel{eq:Sab1} \ee 
where $ \sqrt{-R_4(u,b)} $ was introduced to keep everything real. 
That feature was overlooked in~\cite{Laporta:2004rb}, where however 
the roots $ (0,4m^2) $ were not of interest. \par 

\bibliographystyle{bibliostyle} 
\bibliography{Biblio} 

\providecommand{\href}[2]{#2}\begingroup\raggedright\begin{thebibliography}{10}

\bibitem{Goncharov}
A.~B. Goncharov, {\it {Geometry of configurations, polylogarithms, and motivic
  cohomology}},  {\em Adv. Math.} {\bf 114} (1995), no.~2 197--318.

\bibitem{Remiddi:1999ew}
E.~Remiddi and J.~Vermaseren, {\it {Harmonic polylogarithms}},  {\em
  Int.J.Mod.Phys.} {\bf A15} (2000) 725--754,
  [\href{http://arxiv.org/abs/hep-ph/9905237}{{\tt hep-ph/9905237}}].

\bibitem{Gehrmann:2000zt}
T.~Gehrmann and E.~Remiddi, {\it {Two loop master integrals for $\gamma^*
  \rightarrow$ 3 jets: The Planar topologies}},  {\em Nucl.Phys.} {\bf B601}
  (2001) 248--286, [\href{http://arxiv.org/abs/hep-ph/0008287}{{\tt
  hep-ph/0008287}}].

\bibitem{Vollinga:2004sn}
J.~Vollinga and S.~Weinzierl, {\it {Numerical evaluation of multiple
  polylogarithms}},  {\em Comput.Phys.Commun.} {\bf 167} (2005) 177,
  [\href{http://arxiv.org/abs/hep-ph/0410259}{{\tt hep-ph/0410259}}].

\bibitem{Broadhurst:1987ei}
D.~J. Broadhurst, {\it {The Master Two Loop Diagram With Masses}},  {\em Z.
  Phys.} {\bf C47} (1990) 115--124.

\bibitem{Bauberger:1994by}
S.~Bauberger, F.~A. Berends, M.~Bohm, and M.~Buza, {\it {Analytical and
  numerical methods for massive two loop selfenergy diagrams}},  {\em Nucl.
  Phys.} {\bf B434} (1995) 383--407,
  [\href{http://arxiv.org/abs/hep-ph/9409388}{{\tt hep-ph/9409388}}].

\bibitem{Bauberger:1994hx}
S.~Bauberger and M.~Bohm, {\it {Simple one-dimensional integral representations
  for two loop selfenergies: The Master diagram}},  {\em Nucl. Phys.} {\bf
  B445} (1995) 25--48, [\href{http://arxiv.org/abs/hep-ph/9501201}{{\tt
  hep-ph/9501201}}].

\bibitem{Caffo:1998du}
M.~Caffo, H.~Czyz, S.~Laporta, and E.~Remiddi, {\it {The Master differential
  equations for the two loop sunrise selfmass amplitudes}},  {\em Nuovo Cim.}
  {\bf A111} (1998) 365--389, [\href{http://arxiv.org/abs/hep-th/9805118}{{\tt
  hep-th/9805118}}].

\bibitem{Laporta:2004rb}
S.~Laporta and E.~Remiddi, {\it {Analytic treatment of the two loop equal mass
  sunrise graph}},  {\em Nucl.Phys.} {\bf B704} (2005) 349--386,
  [\href{http://arxiv.org/abs/hep-ph/0406160}{{\tt hep-ph/0406160}}].

\bibitem{Bloch:2013tra}
S.~Bloch and P.~Vanhove, {\it {The elliptic dilogarithm for the sunset graph}},
   \href{http://arxiv.org/abs/1309.5865}{{\tt arXiv:1309.5865}}.

\bibitem{Remiddi:2013joa}
E.~Remiddi and L.~Tancredi, {\it {Schouten identities for Feynman graph
  amplitudes; The Master Integrals for the two-loop massive sunrise graph}},
  {\em Nucl.Phys.} {\bf B880} (2014) 343--377,
  [\href{http://arxiv.org/abs/1311.3342}{{\tt arXiv:1311.3342}}].

\bibitem{Adams:2013nia}
L.~Adams, C.~Bogner, and S.~Weinzierl, {\it {The two-loop sunrise graph with
  arbitrary masses}},  {\em J.Math.Phys.} {\bf 54} (2013) 052303,
  [\href{http://arxiv.org/abs/1302.7004}{{\tt arXiv:1302.7004}}].

\bibitem{Adams:2014vja}
L.~Adams, C.~Bogner, and S.~Weinzierl, {\it {The two-loop sunrise graph in two
  space-time dimensions with arbitrary masses in terms of elliptic
  dilogarithms}},  {\em J.Math.Phys.} {\bf 55} (2014), no.~10 102301,
  [\href{http://arxiv.org/abs/1405.5640}{{\tt arXiv:1405.5640}}].

\bibitem{Adams:2015gva}
L.~Adams, C.~Bogner, and S.~Weinzierl, {\it {The two-loop sunrise integral
  around four space-time dimensions and generalisations of the Clausen and
  Glaisher functions towards the elliptic case}},
  \href{http://arxiv.org/abs/1504.03255}{{\tt arXiv:1504.03255}}.

\bibitem{Adams:2015ydq}
L.~Adams, C.~Bogner, and S.~Weinzierl, {\it {The iterated structure of the
  all-order result for the two-loop sunrise integral}},
  \href{http://arxiv.org/abs/1512.05630}{{\tt arXiv:1512.05630}}.

\bibitem{Remiddi:2016gno}
E.~Remiddi and L.~Tancredi, {\it {Differential equations and dispersion
  relations for Feynman amplitudes. The two-loop massive sunrise and the kite
  integral}},  {\em Nucl. Phys.} {\bf B907} (2016) 400--444,
  [\href{http://arxiv.org/abs/1602.01481}{{\tt arXiv:1602.01481}}].

\bibitem{Adams:2016xah}
L.~Adams, C.~Bogner, A.~Schweitzer, and S.~Weinzierl, {\it {The kite integral
  to all orders in terms of elliptic polylogarithms}},  {\em J. Math. Phys.}
  {\bf 57} (2016) 122302, [\href{http://arxiv.org/abs/1607.01571}{{\tt
  arXiv:1607.01571}}].

\bibitem{Bonciani:2016qxi}
R.~Bonciani, V.~Del~Duca, H.~Frellesvig, J.~M. Henn, F.~Moriello, and V.~A.
  Smirnov, {\it {Two-loop planar master integrals for Higgs$\to 3$ partons with
  full heavy-quark mass dependence}},  {\em JHEP} {\bf 12} (2016) 096,
  [\href{http://arxiv.org/abs/1609.06685}{{\tt arXiv:1609.06685}}].

\bibitem{Bloch:2016izu}
S.~Bloch, M.~Kerr, and P.~Vanhove, {\it {Local mirror symmetry and the sunset
  Feynman integral}},  \href{http://arxiv.org/abs/1601.08181}{{\tt
  arXiv:1601.08181}}.

\bibitem{Passarino:2016zcd}
G.~Passarino, {\it {Elliptic Polylogarithms and Basic Hypergeometric
  Functions}},  {\em Eur. Phys. J.} {\bf C77} (2017), no.~2 77,
  [\href{http://arxiv.org/abs/1610.06207}{{\tt arXiv:1610.06207}}].

\bibitem{vonManteuffel:2017hms}
A.~von Manteuffel and L.~Tancredi, {\it {A non-planar two-loop three-point
  function beyond multiple polylogarithms}},  {\em JHEP} {\bf 06} (2017) 127,
  [\href{http://arxiv.org/abs/1701.05905}{{\tt arXiv:1701.05905}}].

\bibitem{Adams:2017ejb}
L.~Adams and S.~Weinzierl, {\it {Feynman integrals and iterated integrals of
  modular forms}},  \href{http://arxiv.org/abs/1704.08895}{{\tt
  arXiv:1704.08895}}.

\bibitem{Ablinger:2017bjx}
J.~Ablinger, J.~Bluemlein, A.~De~Freitas, M.~van Hoeij, E.~Imamoglu, C.~G.
  Raab, C.~S. Radu, and C.~Schneider, {\it {Iterated Elliptic and
  Hypergeometric Integrals for Feynman Diagrams}},
  \href{http://arxiv.org/abs/1706.01299}{{\tt arXiv:1706.01299}}.

\bibitem{Brown:MEPLs}
F.~Brown and A.~Levin, {\it {Multiple Elliptic Polylogarithms}},
  \href{http://arxiv.org/abs/1110.6917}{{\tt arXiv:1110.6917}}.

\bibitem{Broedel:2015hia}
J.~Broedel, N.~Matthes, and O.~Schlotterer, {\it {Relations between elliptic
  multiple zeta values and a special derivation algebra}},
  \href{http://arxiv.org/abs/1507.02254}{{\tt arXiv:1507.02254}}.

\bibitem{Sabry}
A.~Sabry, {\it {Fourth order spectral functions for the electron propagator}},
  {\em Nucl. Phys.} {\bf 33} (1962), no.~17 401--430.

\bibitem{Laporta:2001dd}
S.~Laporta, {\it {High precision calculation of multiloop Feynman integrals by
  difference equations}},  {\em Int.J.Mod.Phys.} {\bf A15} (2000) 5087--5159,
  [\href{http://arxiv.org/abs/hep-ph/0102033}{{\tt hep-ph/0102033}}].

\bibitem{Primo:2016ebd}
A.~Primo and L.~Tancredi, {\it {On the maximal cut of Feynman integrals and the
  solution of their differential equations}},  {\em Nucl. Phys.} {\bf B916}
  (2017) 94--116, [\href{http://arxiv.org/abs/1610.08397}{{\tt
  arXiv:1610.08397}}].

\bibitem{Frellesvig:2017aai}
H.~Frellesvig and C.~G. Papadopoulos, {\it {Cuts of Feynman Integrals in Baikov
  representation}},  {\em JHEP} {\bf 04} (2017) 083,
  [\href{http://arxiv.org/abs/1701.07356}{{\tt arXiv:1701.07356}}].

\bibitem{Bosma:2017ens}
J.~Bosma, M.~Sogaard, and Y.~Zhang, {\it {Maximal Cuts in Arbitrary
  Dimension}},  {\em JHEP} {\bf 08} (2017) 051,
  [\href{http://arxiv.org/abs/1704.04255}{{\tt arXiv:1704.04255}}].

\bibitem{Primo:2017ipr}
A.~Primo and L.~Tancredi, {\it {Maximal cuts and differential equations for
  Feynman integrals. An application to the three-loop massive banana graph}},
  {\em Nucl. Phys.} {\bf B921} (2017) 316--356,
  [\href{http://arxiv.org/abs/1704.05465}{{\tt arXiv:1704.05465}}].

\bibitem{Harley:2017qut}
M.~Harley, F.~Moriello, and R.~M. Schabinger, {\it {Baikov-Lee Representations
  Of Cut Feynman Integrals}},  {\em JHEP} {\bf 06} (2017) 049,
  [\href{http://arxiv.org/abs/1705.03478}{{\tt arXiv:1705.03478}}].

\bibitem{CaronHuot:2012ab}
S.~Caron-Huot and K.~J. Larsen, {\it {Uniqueness of two-loop master contours}},
   {\em JHEP} {\bf 10} (2012) 026, [\href{http://arxiv.org/abs/1205.0801}{{\tt
  arXiv:1205.0801}}].

\bibitem{Huang:2013kh}
R.~Huang and Y.~Zhang, {\it {On Genera of Curves from High-loop Generalized
  Unitarity Cuts}},  {\em JHEP} {\bf 04} (2013) 080,
  [\href{http://arxiv.org/abs/1302.1023}{{\tt arXiv:1302.1023}}].

\bibitem{Sogaard:2014ila}
M.~Sogaard and Y.~Zhang, {\it {Unitarity Cuts of Integrals with Doubled
  Propagators}},  {\em JHEP} {\bf 07} (2014) 112,
  [\href{http://arxiv.org/abs/1403.2463}{{\tt arXiv:1403.2463}}].

\bibitem{Hauenstein:2014mda}
J.~D. Hauenstein, R.~Huang, D.~Mehta, and Y.~Zhang, {\it {Global Structure of
  Curves from Generalized Unitarity Cut of Three-loop Diagrams}},  {\em JHEP}
  {\bf 02} (2015) 136, [\href{http://arxiv.org/abs/1408.3355}{{\tt
  arXiv:1408.3355}}].

\bibitem{Abreu:2017ptx}
S.~Abreu, R.~Britto, C.~Duhr, and E.~Gardi, {\it {Cuts from residues: the
  one-loop case}},  {\em JHEP} {\bf 06} (2017) 114,
  [\href{http://arxiv.org/abs/1702.03163}{{\tt arXiv:1702.03163}}].

\bibitem{Abreu:2017enx}
S.~Abreu, R.~Britto, C.~Duhr, and E.~Gardi, {\it {Algebraic Structure of Cut
  Feynman Integrals and the Diagrammatic Coaction}},  {\em Phys. Rev. Lett.}
  {\bf 119} (2017), no.~5 051601, [\href{http://arxiv.org/abs/1703.05064}{{\tt
  arXiv:1703.05064}}].

\bibitem{Abreu:2017mtm}
S.~Abreu, R.~Britto, C.~Duhr, and E.~Gardi, {\it {Diagrammatic Hopf algebra of
  cut Feynman integrals: the one-loop case}},
  \href{http://arxiv.org/abs/1704.07931}{{\tt arXiv:1704.07931}}.

\bibitem{Vermaseren:2000nd}
J.~Vermaseren, {\it {New features of FORM}},
  \href{http://arxiv.org/abs/math-ph/0010025}{{\tt math-ph/0010025}}.

\bibitem{Barbieri:1974nc}
R.~Barbieri and E.~Remiddi, {\it {Electron and Muon 1/2(g-2) from Vacuum
  Polarization Insertions}},  {\em Nucl. Phys.} {\bf B90} (1975) 233--266.

\end{thebibliography}\endgroup
\end{document}